\documentclass[12pt,letterpaper]{article}
\usepackage[T1]{fontenc}
\usepackage{amsmath,amssymb}
\usepackage{graphicx}
\usepackage{placeins}
\usepackage{booktabs,tabularx,array}
\usepackage{newtxtext,newtxmath}
\usepackage[top=30mm,bottom=30mm,left=25mm,right=25mm]{geometry}
\usepackage{cite}
\usepackage{hyperref}
\hypersetup{pdftitle={Analytic Bounds on Conformal Defect Spectra},pdfauthor={Yuma Furuta; Wataru Harada; Yuya Kusuki; Yin Tang},pdfsubject={JHEP manuscript draft}}
\numberwithin{equation}{section}
\newcommand{\id}{\mathbf 1}

\begin{document}
\hypersetup{pageanchor=false}
\begin{titlepage}
\thispagestyle{empty}
\newgeometry{left=24.2mm,textwidth=160mm,top=33.4mm,textheight=220mm}
\begin{flushright}
\small
KYUSHU-HET-382\\
RIKEN-iTHEMS-Report-26
\end{flushright}
\vspace{10mm}
\begin{center}
{\large\bfseries Analytic Bounds on Conformal Defect Spectra\par}
\vspace{9mm}
{\normalsize Yuma Furuta\textsuperscript{1}, Wataru Harada\textsuperscript{2}, Yuya Kusuki\textsuperscript{1,2,3,4}, Yin Tang\textsuperscript{1} \par}
\vspace{5mm}
{\small\itshape
\textsuperscript{1}Institute for Advanced Study, Kyushu University, Fukuoka 812-8581, Japan\par
\vspace{2mm}
\textsuperscript{2}Department of Physics, Kyushu University, Fukuoka 819-0395, Japan\par
\vspace{2mm}
\textsuperscript{3}RIKEN Center for Interdisciplinary Theoretical and Mathematical Sciences (iTHEMS), Wako, Saitama 351-0198, Japan\par
\vspace{2mm}
\textsuperscript{4}JST, PRESTO, 4-1-8 Honcho, Kawaguchi, Saitama 332-0012, Japan\par
}
\end{center}
\vspace{10mm}
\begin{abstract}
Conformal defects describe impurities and boundaries at criticality. Their stability depends on the scaling dimensions of operators localized on the defects, but these dimensions are often difficult to determine. We derive analytic upper bounds on the lowest nonvacuum scaling dimension using only two defect entropies and the dimension of a defect-changing operator. The bounds identify when relevant operators must appear on the defect with larger defect entropy. They follow from the conformal bootstrap of four-point functions of defect-changing operators and apply to unitary conformal line defects and to conformal boundaries in two-dimensional CFTs. The Ising and Potts models provide exact checks. In three-dimensional $O(N)$ models, we apply the bounds to direct sums of pinning field defects with symmetry-related orientations. Data from a single such defect exclude stable impurity order with three or more orientations in $O(2)$ and $O(3)$, or four or more in $O(4)$, for any geometry of the orientations, and require a relevant domain-wall operator between individually stable pinning field defects at every nonzero relative angle up to $120^\circ$ in $O(3)$ and up to $109.47^\circ$ in $O(4)$. A direct fuzzy-sphere computation of the lowest domain-wall dimension at different relative angles for $N=2,3,4$ is consistent with these bounds and indicates that, within this class of impurities, only two-state order remains a candidate for stable order in $O(3)$ and $O(4)$, while none survives in $O(2)$. A separate application to AdS/BCFT shows that, at fixed unequal brane tensions, a light boundary-condition-changing operator requires additional light primaries on the boundary with larger boundary entropy.
\end{abstract}
\end{titlepage}
\restoregeometry
\hypersetup{pageanchor=true}
\tableofcontents
\clearpage

\section{Introduction \& Summary}
Impurities and boundaries can support critical behavior distinct from that of the surrounding bulk~\cite{CardyReview,Billo}. Whether an impurity fixed point is stable depends on the scaling dimensions of operators localized at the impurity, but these dimensions can be difficult to determine even when other defect observables are known. We show that two defect entropies and the dimension of a defect-changing operator already bound the lowest nonvacuum scaling dimension on the defect with larger defect entropy. These few data can therefore rule out a proposed stable fixed point without calculating its local spectrum. For an impurity with several degenerate ordered configurations, the same bounds constrain how many configurations can remain stable against symmetry-preserving perturbations.

Our starting point is the possibility of bounding the gap in one crossing channel without assuming a gap in the other~\cite{HMR}. Defect-changing operators, inserted where one defect condition changes into another, provide such a crossing problem: the two channels of their alternating four-point function describe local operators on different defects. Crossing symmetry and Ward identities have been used to constrain boundary and defect spectra both numerically and analytically~\cite{LiendoRastelliVanRees,GaiottoMazacPaulos,GimenezGrauEtAl,BianchiMagnetic,DeyGhosh,GiraultPaulosVanVliet}. In particular, a numerical study of four-point functions of endpoint operators between a line defect and the trivial defect obtained bounds on defect $g$-functions and on the dimensions of endpoint and domain-wall operators~\cite{LLM}. Combined with spectral data for the three-dimensional Ising bulk, these bounds provided evidence that impurity order with two degenerate configurations is unstable. Our analytic bounds require no detailed information about the bulk spectrum, and their explicit dependence on defect entropy allows us to use the same data to test the stability of different impurity configurations. The exact boundary spectra of the Ising and Potts models, including stable tricritical examples, provide checks in Section~\ref{sec:examples}.

To study impurity order, we use a direct sum of symmetry-related pinning field defects, with each defect representing one degenerate ordered configuration. Defect-changing operators between these defects, called domain-wall operators, can generate symmetry-preserving perturbations that destabilize the order even when the individual defects are stable~\cite{LLM,Yang}. In three-dimensional $O(N)$ models, numerical studies of domain-wall operators between defects with opposite orientations support instability of two-state order for $N=1,2$ and irrelevance of the domain-wall perturbation for $N=3,4$~\cite{Yang}. These results do not determine whether impurity order with more than two configurations is stable. For three pinning field defects with orientations separated by $120^\circ$ in an internal plane, stability depends on the dimension of the corresponding domain-wall operator. Computing this dimension was proposed in Ref.~\cite{Yang}. The dimensions of domain-wall operators at arbitrary internal angles are known in the $4-\epsilon$ expansion~\cite{CuomoHeKomargodski}, but extrapolation to three dimensions does not by itself determine stability with controlled errors. We compute this dimension, and the dimensions of domain-wall operators at other relative angles, directly on the fuzzy sphere in Section~\ref{sec:numerical}.

Table~\ref{tab:prior_comparison} compares these results and their consequences with our analytic bounds and with our direct computation. With the reported numerical inputs and the assumptions of Section~\ref{sec:o3}, the bounds exclude stable impurity order for every symmetric configuration of three or more orientations in $O(2)$ and $O(3)$ and of four or more in $O(4)$, whatever the geometry of the orientations. This constrains the number of ordered configurations even when numerical results are consistent with stable two-state order.

\begin{table}[!htb]
\centering
\small
\setlength{\tabcolsep}{4pt}
\renewcommand{\arraystretch}{1.13}
\begin{tabularx}{\textwidth}{@{}lc>{\raggedright\arraybackslash}X>{\raggedright\arraybackslash}p{3.6cm}>{\raggedright\arraybackslash}p{3.0cm}@{}}
\toprule
Bulk CFT & $q$ & Earlier results and their consequences & Present analytic bounds & Direct computation \\
\midrule
Ising & $2$ & Instability supported by calculations of the dimensions of domain-wall operators and the bootstrap of endpoint operators~\cite{LLM,ZhouGaiottoHeZou,CuomoHeKomargodski}. & No exclusion. & --- \\
\midrule
$O(2)$ & $2$ & The domain-wall operator between opposite orientations is relevant in the numerical analysis~\cite{Yang}. & No exclusion. & $\Delta_{\rm DW}(180^\circ)\simeq0.85$--$0.86$: relevant. \\
 & $\geq3$ & For regular planar configurations with even $q$, instability follows from the result for opposite orientations. That argument does not cover odd $q$ or other symmetric geometries. & All $q\geq3$ excluded for any symmetric geometry; $H_a\leq0.773$ at $q=3$. & $\Delta_{\rm DW}(120^\circ)\simeq0.50$--$0.56$: relevant. \\
\midrule
$O(3)$ & $2$ & Numerical extrapolation indicates that the domain-wall operator between opposite orientations is irrelevant~\cite{Yang}. & No exclusion. & $\Delta_{\rm DW}(180^\circ)\simeq1.00$--$1.15$: near threshold. \\
 & $\geq3$ & Computing the dimension of the domain-wall operator at $120^\circ$ was proposed to determine the stability of three-state order~\cite{Yang}. & All $q\geq3$ excluded for any symmetric geometry; $H_a\leq0.987$ at $q=3$; $\Delta_{\rm DW}(\theta)<1$ for $0<\theta\leq120^\circ$. & $\Delta_{\rm DW}(120^\circ)\simeq0.65$--$0.67$: relevant. \\
\midrule
$O(4)$ & $2$ & Numerical extrapolation indicates that the domain-wall operator between opposite orientations is irrelevant~\cite{Yang}. & No exclusion. & $\Delta_{\rm DW}(180^\circ)\simeq1.17$--$1.42$: irrelevant. \\
 & $3$ & Not determined by the data for opposite orientations. & No exclusion. & $\Delta_{\rm DW}(120^\circ)\simeq0.77$--$0.84$: relevant. \\
 & $\geq4$ & Data for opposite orientations alone do not determine the dimensions of domain-wall operators between other orientations. & All $q\geq4$ excluded for any symmetric geometry; $H_a\leq0.946$ at $q=4$; $\Delta_{\rm DW}(\theta)<1$ for $0<\theta\leq109.47^\circ$. & $\Delta_{\rm DW}(109.47^\circ)\simeq0.66$--$0.70$: relevant. \\
\bottomrule
\end{tabularx}
\caption{Earlier results, analytic constraints, and fuzzy-sphere estimates for ordered impurities with $q$ symmetry-related pinning orientations (opposite orientations for Ising); perturbations preserve the symmetry permuting them. $\Delta_{\rm DW}(\theta)$ is the lowest dimension of a scalar domain-wall operator between two pinning field defects at internal angle $\theta$, and $H_a$ the lowest nonvacuum dimension in the OPE of the invariant endpoint operator with its adjoint. Bounds use the inputs and assumptions of Section~\ref{sec:o3} and are rounded upward; the last column is the fit spread of Table~\ref{tab:dwnumerics}. \emph{No exclusion} does not imply stability.}
\label{tab:prior_comparison}
\end{table}

The input needed to apply these bounds is the defect entropy of a single pinning field defect and the dimension of its endpoint operator. Using the same data, we can therefore test the stability of different arrangements without separately calculating the lowest dimension of a domain-wall operator for each pair, and the instability criterion for a symmetric direct sum turns out not to depend on the geometry of its orientations. Configurations in which all pairs of orientations make a common angle, obtained by tilting a regular simplex, bound the dimension of the domain-wall operator over continuous ranges of angles: if the individual pinning field defects are stable, a relevant domain-wall operator must exist at every nonzero relative angle up to $120^\circ$ in $O(3)$ and up to $109.47^\circ$ in $O(4)$, without interpolation between angles or an assumption that this dimension varies monotonically with the angle.

To test these conclusions, we compute the lowest domain-wall dimension at a grid of relative angles for the $O(2)$, $O(3)$, and $O(4)$ models by exact diagonalization on the fuzzy sphere, with up to twelve orbitals (Section~\ref{sec:numerical}). The estimates lie below every applicable bound; for the three principal exclusions, the fitted continuum estimates lie approximately $25$--$35\%$ below the corresponding bounds at the nominal inputs. They support the exclusions without using the defect entropy or the endpoint dimension. They also indicate that the domain-wall operator between opposite orientations is irrelevant in $O(4)$ and close to the relevance threshold in $O(3)$, so that within this class of impurities only two-state order remains a candidate for $N=3,4$, while the estimates support the absence of stable order for $N=2$, subject to the symmetry and extrapolation assumptions stated in Section~\ref{sec:physics}. Section~\ref{sec:physics} also discusses why the bounds carry information not implied by simpler considerations---exact boundary spectra of two-dimensional parafermion theories show that the domain-wall dimension need not increase with the relative angle---and how the growth of the domain-wall dimensions with $N$ reflects the increasing rigidity of the pinned orientation.

A separate application tests boundary spectra in AdS$_3$/BCFT$_2$, where conformal boundaries are represented by end-of-the-world branes. A junction between two branes corresponds to a boundary-condition-changing operator~\cite{MiyajiMurdia,BiswasEtAl}. At fixed unequal brane tensions, a light boundary-condition-changing operator requires additional light primary operators on the boundary with larger boundary entropy. Here light means that the dimension grows more slowly than the central charge. For the brane solutions of Ref.~\cite{MiyajiMurdia}, our bounds test whether their junction geometries can coexist with an assumed gap in the boundary spectrum. These constraints complement the annulus bootstrap~\cite{CMW}, which equates the open- and closed-channel expansions of the annulus partition function.
\FloatBarrier

\section{Four-point Functions of Defect-changing Operators}
\label{sec:crossing}
\begin{figure}[!t]
\centering
\includegraphics[width=0.88\textwidth]{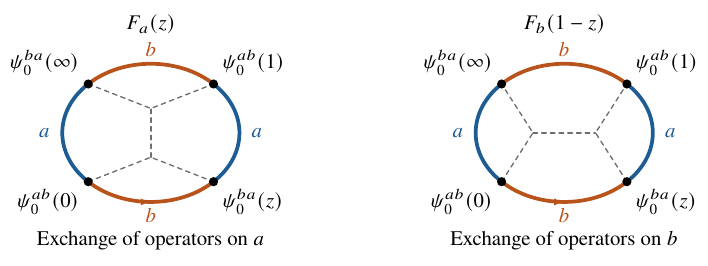}
\caption{The same four-point function admits OPE expansions into operators on $a$ (left) or $b$ (right). Solid segments carry the labeled defect conditions; dashed lines indicate the OPE pairings.}
\label{fig:crossing}
\end{figure}

Consider two conformal line defects $a,b$ in a bulk conformal field theory (CFT), or two conformal boundary conditions in a two-dimensional CFT. We assume unitarity throughout. We denote defect-changing primaries of scaling dimension $h_i^{ab}$ by $\psi_i^{ab}$, with adjoints $\psi_i^{ba}$. The superscripts specify the defects on the two sides of the insertion; $\psi_i^{aa}$ are local primaries on $a$. For conformal boundaries, $\psi_i^{ab}$ are boundary-condition-changing operators.

The defect $g$-function $g_a$ is the expectation value of the defect on a circle, normalized by the defect-free bulk partition function. For a conformal boundary, $g_a$ is the overlap of the boundary state with the bulk vacuum. The defect entropy is $\log g_a$~\cite{CardyReview,AffleckLudwig}. We assume that $g_a$ and $g_b$ are finite and positive. The vacuum primary $\psi_0^{aa}=\id_a$ acts as the unit on $a$, with $\langle\id_a\rangle=g_a$. We use the two-point normalization
\begin{equation}
\langle\psi_i^{ab}(0)\psi_j^{ba}(\infty)\rangle
=\sqrt{g_ag_b}\,\delta_{ij}.
\label{eq:twopointnormalization}
\end{equation}
We choose an external primary $\psi_0^{ab}$ of dimension $\delta=h_0^{ab}>0$ and its adjoint $\psi_0^{ba}$. For line defects, we take these operators to be scalars under transverse rotations. In the $a$-channel operator product expansion (OPE) of this pair, we assume that the vacuum is the only primary with $h_i^{aa}=0$ that contributes. We impose the same condition for $h_i^{bb}=0$ in the $b$ channel. Other primaries with these dimensions may exist, but their coefficients in these OPEs must vanish. Section~\ref{sec:stability} constructs direct sums satisfying this condition.

Place alternating insertions of $\psi_0^{ab}$ and $\psi_0^{ba}$ at $0,z,1,\infty$, where $z$ is the cross ratio and $0<z<1$ (see Fig.~\ref{fig:crossing}). We write the OPE with the dependence on the defect $g$-functions explicit:
\begin{equation}
\psi_0^{ab}(0)\psi_0^{ba}(z)=\sqrt{\frac{g_b}{g_a}}\sum_i C_i^{(a)}z^{h_i^{aa}-2\delta}\bigl[\psi_i^{aa}(0)+\text{descendants}\bigr].
\label{eq:opecoefficients}
\end{equation}
The OPE coefficient of the vacuum, $C_0^{(a)}=1$, follows from Eq.~\eqref{eq:twopointnormalization}. Let $\mathcal{F}_h(z)$ denote the conformal block of an intermediate primary of dimension $h$. The conformal block expansion in the $a$ channel and the crossing equation are~\cite{Lewellen,LLM}
\begin{gather}
F_a(z)=g_b\left[\mathcal{F}_0(z)+\sum_{h>0}p_h^{(a)}\mathcal{F}_h(z)\right],\qquad p_h^{(a)}\geq0,\label{eq:blocks}\\
F_a(z)=\left(\frac{z}{1-z}\right)^{2\delta}F_b(1-z).\label{eq:crossing}
\end{gather}
Here $F_a(z)$ is defined by
\begin{equation}
\langle\psi_0^{ab}(0)\psi_0^{ba}(z)\psi_0^{ab}(1)\psi_0^{ba}(\infty)\rangle
=\frac{F_a(z)}{z^{2\delta}},
\label{eq:fourpoint}
\end{equation}
and $p_h^{(a)}$ is the sum of squared OPE coefficients of primaries with dimension $h$:
\begin{equation}
p_h^{(a)}:=\sum_{i:\,h_i^{aa}=h}|C_i^{(a)}|^2.
\label{eq:opeweights}
\end{equation}
We define $F_b(z)$ and $p_h^{(b)}$ by interchanging $a$ and $b$.

For a general line defect, we use the global conformal blocks
\begin{equation}
\mathcal{F}_h(z)=z^h{}_2F_1(h,h;2h;z),\qquad h>0,\qquad \mathcal{F}_0(z)=1.
\label{eq:globalblock}
\end{equation}
Here ${}_2F_1$ is the hypergeometric function. In a two-dimensional boundary CFT, we instead use Virasoro blocks with bulk central charge $c$ and external dimension $\delta$. The Virasoro vacuum block includes the entire vacuum representation. Sums over a continuous spectrum are understood as integrals.

For a given intermediate dimension $h$, the conformal blocks have the same functional form in both channels, although the spectra and OPE coefficients can differ. The vacuum contributions in $F_a$ and $F_b$ are multiplied by $g_b$ and $g_a$, respectively. This relative normalization is fixed by the defect entropies; it is equal to one in the crossing problem of Ref.~\cite{HMR}. At $z=1/2$, crossing symmetry and the nonnegative terms in the $b$-channel expansion imply
\begin{equation}
F_a(1/2)=F_b(1/2)\geq g_a\mathcal{F}_0(1/2).
\label{eq:vacuumlower}
\end{equation}
Using this estimate to bound the vacuum contribution on $a$ gives
\begin{equation}
F_a(1/2)-g_b\mathcal{F}_0(1/2)
\geq\left(1-\frac{g_b}{g_a}\right)F_a(1/2).
\label{eq:weight}
\end{equation}
The left-hand side is the nonvacuum contribution on $a$. For $g_a>g_b$, its ratio to $F_a(1/2)$ is therefore bounded below by $1-g_b/g_a$, independently of the spectrum on $b$.

\section{Bound on the Lowest Dimension of a Defect Operator}
\label{sec:first}
Taking the first derivative of the crossing equation gives an upper bound on the lowest scaling dimension in one channel, without assuming a gap in the other. Let $H_a$ denote the lowest dimension of a nonvacuum primary appearing in the $a$-channel expansion~\eqref{eq:blocks}, or the infimum if no minimum is attained. The bound is
\begin{equation}
\left(1-\frac{g_b}{g_a}\right)H_a\leq2\sqrt2\,\delta,\qquad g_a>g_b.
\label{eq:main}
\end{equation}
At fixed $g_a/g_b>1$, the lowest scaling dimension in the chosen OPE on $a$ must therefore tend to zero as $\delta\to0$.

The proof uses the following estimate for global and Virasoro blocks, with primes denoting derivatives with respect to $z$:
\begin{equation}
\mathcal{F}_h'(1/2)\geq2\sqrt2\,h\mathcal{F}_h(1/2),\qquad \mathcal{F}_0'(1/2)\geq0.
\label{eq:lemma}
\end{equation}
For global blocks this follows from their radial expansion with nonnegative coefficients~\cite{HogervorstRychkov}. A Virasoro block in a unitary theory also decomposes into global blocks with nonnegative coefficients. The global primaries in this decomposition have dimensions at least $h$, so the same estimate applies. Appendix~\ref{app:vir} gives the details. Every nonvacuum term in the $a$-channel expansion has $h\geq H_a$, while the derivative of the vacuum block is nonnegative. Applying Eq.~\eqref{eq:lemma} term by term and then using Eq.~\eqref{eq:weight} gives
\begin{equation}
\begin{aligned}
F_a'(1/2)&\geq2\sqrt2\,H_a\left[F_a(1/2)-g_b\mathcal{F}_0(1/2)\right]\\
&\geq2\sqrt2\,H_a\left(1-\frac{g_b}{g_a}\right)F_a(1/2).
\end{aligned}
\label{eq:derivativebound}
\end{equation}
The crossing equation also gives an upper bound on this derivative. Differentiating Eq.~\eqref{eq:crossing} at $z=1/2$ gives
\begin{equation}
F_a'(1/2)+F_b'(1/2)=8\delta F_a(1/2).
\label{eq:derivative}
\end{equation}
Since $F_b'(1/2)\geq0$ by Eq.~\eqref{eq:lemma}, combining Eqs.~\eqref{eq:derivativebound} and \eqref{eq:derivative} proves Eq.~\eqref{eq:main}.

The same argument also constrains the contribution of primaries of small dimension to the four-point function. To see this, retain the individual dimensions $h$ in Eq.~\eqref{eq:lemma} instead of replacing them by $H_a$. Combining this estimate with Eq.~\eqref{eq:derivative} gives
\begin{equation}
g_b\sum_{h>0}h\,p_h^{(a)}\mathcal{F}_h(1/2)\leq2\sqrt2\,\delta F_a(1/2).
\label{eq:moment}
\end{equation}
Split the sum over nonvacuum primaries at a dimension equal to twice the upper bound on $H_a$ in Eq.~\eqref{eq:main}. Since every term in Eq.~\eqref{eq:moment} is nonnegative, the inequality still holds when the sum is restricted to primaries at or above this threshold. Replacing $h$ by the threshold and dividing by it then shows that these primaries contribute at most half the lower bound in Eq.~\eqref{eq:weight} on the total nonvacuum contribution. Subtracting this upper bound from the lower bound on the total nonvacuum contribution gives
\begin{equation}
g_b\sum_{0<h<4\sqrt2\,\delta/(1-g_b/g_a)}p_h^{(a)}\mathcal{F}_h(1/2)
\geq\frac12\left(1-\frac{g_b}{g_a}\right)F_a(1/2).
\label{eq:lightweight}
\end{equation}
At fixed $g_a/g_b>1$, the upper limit on $h$ in the sum tends to zero with $\delta$, while the contribution of these primaries relative to $F_a(1/2)$ remains bounded away from zero.

\section{Bounds from the Third Derivative}
\label{sec:third}
The argument of Section~\ref{sec:first} can also be applied to derivatives of the crossing equation up to third order. With the same defect data, this gives a stronger criterion for the existence of relevant operators. For $g_a>g_b$ and $0<\delta\leq1/6$,
\begin{equation}
\left(1-\frac{g_b}{g_a}\right)\min\{H_a,H_a^3\}
\leq\frac{7\delta}{4(1-2\delta)}.
\label{eq:thirdgap}
\end{equation}
If $H_a\geq1$, Eq.~\eqref{eq:thirdgap} requires
\begin{equation}
(4-15\delta)g_a\leq(4-8\delta)g_b.
\label{eq:third}
\end{equation}
If Eq.~\eqref{eq:third} is violated, a relevant operator must be present. This condition is stronger than the one obtained from Eq.~\eqref{eq:main} in the stated range of $\delta$, since $7/[4(1-2\delta)]<2\sqrt2$. In particular, it detects a relevant operator in the $O(3)$ application of Section~\ref{sec:o3}, where Eq.~\eqref{eq:main} is inconclusive.

To prove Eq.~\eqref{eq:thirdgap}, we again compare a lower bound from the spectrum on $a$ with an upper bound from the crossing equation. We begin with global conformal blocks and evaluate all functions and derivatives at $z=1/2$. The Casimir equation gives
\begin{equation}
\mathcal{F}_h''=2\mathcal{F}_h'+8h(h-1)\mathcal{F}_h,\qquad
\mathcal{F}_h'''=8\bigl[h(h-1)+1\bigr]\mathcal{F}_h'.
\label{eq:casimir}
\end{equation}
Together with the lower bound in Eq.~\eqref{eq:lemma}, these identities imply
\begin{align}
&\mathcal{F}_h'''-6\sqrt2\,\mathcal{F}_h''+12\sqrt2\,\mathcal{F}_h'\nonumber\\
&\qquad=8\bigl[h(h-1)+1\bigr]\mathcal{F}_h'-48\sqrt2\,h(h-1)\mathcal{F}_h\nonumber\\
&\qquad\geq16\sqrt2\,h(h-2)^2\mathcal{F}_h\geq0.
\label{eq:thirdpositive}
\end{align}
The first inequality follows because the coefficient $h(h-1)+1$ is positive, allowing us to apply Eq.~\eqref{eq:lemma} to the first term on the right-hand side.

Differentiating the crossing equation~\eqref{eq:crossing} twice and three times gives
\begin{align}
F_a''-F_b''&=8\delta(F_a'-F_b'),\label{eq:crossingsecond}\\
F_a'''+F_b'''&=24\delta F_b''-192\delta^2 F_b'+64\delta(8\delta^2+1)F_a.
\label{eq:crossingthird}
\end{align}
Combining these identities with $F_a'+F_b'=8\delta F_a$, from Eq.~\eqref{eq:derivative}, yields
\begin{align}
&F_a'''+6(\sqrt2-4\delta)F_a''+12(16\delta^2-8\sqrt2\,\delta+\sqrt2)F_a'\nonumber\\
&\qquad+F_b'''-6\sqrt2\,F_b''+12\sqrt2\,F_b'\nonumber\\
&\qquad=32\delta(16\delta^2-12\sqrt2\,\delta+3\sqrt2+2)F_a.
\label{eq:thirdidentity}
\end{align}

Equation~\eqref{eq:thirdpositive} shows that the three terms involving $F_b$ in Eq.~\eqref{eq:thirdidentity} have a nonnegative sum. Omitting this sum gives an upper bound on the remaining terms. For a lower bound, we apply the Casimir identities~\eqref{eq:casimir} and Eq.~\eqref{eq:lemma} to each conformal block in the expansion of $F_a$:
\begin{align}
&\mathcal{F}_h'''+6(\sqrt2-4\delta)\mathcal{F}_h''+12(16\delta^2-8\sqrt2\,\delta+\sqrt2)\mathcal{F}_h'\nonumber\\
&\qquad\geq16\sqrt2\bigl[24\delta^2-6(2\sqrt2+1)\delta+3\sqrt2+1\bigr]\min\{h,h^3\}\mathcal{F}_h.
\label{eq:thirdlower}
\end{align}
After using the Casimir identities, Eq.~\eqref{eq:lemma} bounds the left-hand side from below by $\mathcal{F}_h$ times a cubic polynomial in $h$. Its coefficients are positive for $0<\delta\leq1/6$, so replacing each of $h$, $h^2$, and $h^3$ by $\min\{h,h^3\}$ gives Eq.~\eqref{eq:thirdlower}. Appendix~\ref{app:third} provides the details.

Summing over the conformal blocks in $F_a$ now converts Eq.~\eqref{eq:thirdlower} into a bound on $H_a$. The global vacuum block contributes zero, while every nonvacuum term has $h\geq H_a$. Since $\min\{h,h^3\}$ increases with $h$, multiplying Eq.~\eqref{eq:thirdlower} by the nonnegative weights $g_bp_h^{(a)}$ and using the upper bound from Eq.~\eqref{eq:thirdidentity} yields
\begin{align}
&16\sqrt2\bigl[24\delta^2-6(2\sqrt2+1)\delta+3\sqrt2+1\bigr]\min\{H_a,H_a^3\}
\left(F_a-g_b\mathcal{F}_0\right)\nonumber\\
&\qquad\leq32\delta(16\delta^2-12\sqrt2\,\delta+3\sqrt2+2)F_a.
\label{eq:thirdupper}
\end{align}
Equation~\eqref{eq:weight} bounds the nonvacuum contribution $F_a-g_b\mathcal{F}_0$ from below. Inserting that bound and dividing by the positive factors gives
\begin{align}
\left(1-\frac{g_b}{g_a}\right)\min\{H_a,H_a^3\}
&\leq\frac{\sqrt2\,\delta(16\delta^2-12\sqrt2\,\delta+3\sqrt2+2)}{24\delta^2-6(2\sqrt2+1)\delta+3\sqrt2+1}\nonumber\\
&\leq\frac{7\delta}{4(1-2\delta)}.
\label{eq:thirdrational}
\end{align}
The final inequality, proved in Appendix~\ref{app:third}, establishes Eq.~\eqref{eq:thirdgap}. The first upper bound in Eq.~\eqref{eq:thirdrational} is stronger than its simplified form, so we retain it in the $O(N)$ applications of Section~\ref{sec:o3}.

The proof extends to Virasoro blocks through their decomposition into global blocks with nonnegative coefficients (Appendix~\ref{app:vir}). For a Virasoro primary of dimension $h$, all global primaries in this decomposition have dimensions at least $h$. Equation~\eqref{eq:thirdpositive} therefore holds term by term, and the monotonicity of $\min\{h,h^3\}$ ensures that Eq.~\eqref{eq:thirdlower} holds for each nonvacuum Virasoro block. The Virasoro vacuum block also gives nonnegative contributions to the left-hand sides of both inequalities. These contributions can be omitted when deriving Eq.~\eqref{eq:thirdupper}, where $F_a-g_b\mathcal{F}_0$ contains only nonvacuum Virasoro blocks.

\section{Spectra and Stability of Direct Sums of Defects}
\label{sec:stability}
An impurity with several degenerate ordered configurations can be described by a direct sum of conformal defects, one for each configuration. The bounds of Sections~\ref{sec:first} and \ref{sec:third} then constrain how many such configurations can remain stable. Consider $q$ inequivalent defects $a_i$ permuted transitively by a finite symmetry group, with a common defect $g$-function $g_{\rm pin}$. We take each $a_i$ to be elementary, meaning that its spectrum contains a unique vacuum state. We also assume that no defect-changing operator with $h=0$ connects $a_i$ and $a_j$ with $i\ne j$. Let $b$ be invariant under the symmetry, and choose primaries $\psi_0^{a_i b}$ of common dimension $\delta$ that are permuted in the same way as the $a_i$. We form the direct sum and the invariant primary
\begin{equation}
a=\bigoplus_{i=1}^{q}a_i,\qquad
g_a=qg_{\rm pin},\qquad
\psi_0^{ab}=q^{-1/4}\sum_{i=1}^{q}\psi_0^{a_i b}.
\label{eq:sum}
\end{equation}
The prefactor ensures the normalization in Eq.~\eqref{eq:twopointnormalization}. The spectrum on $a$ contains both operators on the individual defects $a_i$ and defect-changing operators between $a_i$ and $a_j$ with $i\ne j$. The latter are called domain-wall operators~\cite{LLM}.

In the direct sum $a$, the vacuum operator $\id_{a_i}$ of each defect $a_i$ acts as the projector onto that defect. Because the symmetry permutes these projectors transitively, their sum $\id_a=\sum_i\id_{a_i}$ is the only invariant combination. The OPE of the invariant primary $\psi_0^{ab}$ with its adjoint can therefore contain only $\id_a$ at $h=0$, as required in Section~\ref{sec:crossing}. With the normalization in Eq.~\eqref{eq:twopointnormalization}, its contribution to the four-point function is
\begin{equation}
\frac{g_b}{g_a}\langle\id_a(0)\id_a(\infty)\rangle\mathcal{F}_0(z)=g_b\mathcal{F}_0(z).
\label{eq:sumvacuum}
\end{equation}
In the $b$ channel, the same requirement holds automatically if $b$ is elementary; otherwise we assume that only the vacuum contributes at $h=0$ in the chosen OPE.

The nonvacuum terms in the OPE on $a$ are also invariant under the symmetry. A primary with $0<h<1$ gives a relevant symmetry-preserving perturbation: a Hermitian combination can be integrated along the defect with a coupling of dimension $1-h>0$. Stability against such perturbations therefore requires $H_a\geq1$. Substituting $g_a=qg_{\rm pin}$ into Eq.~\eqref{eq:main} gives the necessary condition
\begin{equation}
q\leq\frac{g_b}{g_{\rm pin}(1-2\sqrt2\,\delta)},\qquad 2\sqrt2\,\delta<1.
\label{eq:qcap}
\end{equation}
For $0<\delta\leq1/6$, Eq.~\eqref{eq:third} gives the stronger condition
\begin{equation}
q\leq\frac{(4-8\delta)g_b}{(4-15\delta)g_{\rm pin}}.
\label{eq:qthird}
\end{equation}
At fixed $g_{\rm pin}$, $g_b$, and $\delta$, these conditions exclude stability for sufficiently large $q$. Using the same defect data, we can thus bound the number of configurations compatible with stability without knowing the dimensions of domain-wall operators.

Suppose that, on every defect $a_i$, the OPE of $\psi_0^{a_i b}$ with its adjoint contains no primary with $0<h<1$. Any relevant primary required by the bounds must then be a domain-wall operator. This also allows us to constrain the spectrum of domain-wall operators between a given pair of defects by including them in a larger direct sum. Adding defects to the direct sum under the same assumptions increases $g_a$ while keeping $\delta$ fixed. This can strengthen the bound on $H_a$ without changing the spectrum between the original pair. If symmetry relates all pairs of distinct defects $a_i$, the spectrum between any pair is the same, so a relevant domain-wall operator required by the bound must occur between every pair.

The construction also gives spectral bounds without a symmetry relating the defects. Take inequivalent elementary defects $a_i$ with possibly different $g_{a_i}$, and assume that each can be connected to the same defect $b$ by a primary $\psi_0^{a_i b}$ of dimension $\delta$. For their direct sum, set
\begin{equation}
g_a=\sum_i g_{a_i},\qquad
\psi_0^{ab}=\sum_i\left(\frac{g_{a_i}}{g_a}\right)^{1/4}\psi_0^{a_i b}.
\label{eq:weightedsum}
\end{equation}
The common dimension ensures that $\psi_0^{ab}$ is a primary. The weights give the normalization in Eq.~\eqref{eq:twopointnormalization} and equal coefficients for the projectors $\id_{a_i}$ in its OPE with its adjoint. These terms again combine into $\id_a$, giving the vacuum contribution in Eq.~\eqref{eq:sumvacuum}. Under the same assumption on the $b$-channel OPE, Eq.~\eqref{eq:third} gives a necessary condition for the absence of relevant primaries in the chosen OPE on $a$:
\begin{equation}
(4-15\delta)\sum_i g_{a_i}\leq(4-8\delta)g_b,
\qquad 0<\delta\leq\frac16.
\label{eq:sumcriterion}
\end{equation}
If the OPE on each $a_i$ contains no relevant primary and Eq.~\eqref{eq:sumcriterion} is violated, a relevant domain-wall operator must be present between at least one pair of distinct defects $a_i$ and $a_j$.

\section{Exact Checks in Two-dimensional Boundary CFT}
\label{sec:examples}
The exact boundary spectra of the critical Ising and three-state Potts models contain the relevant operators required by the bounds. Their tricritical counterparts provide stable ordered boundaries consistent with the bound in Eq.~\eqref{eq:main}. In the Ising model, take $a$ to be the free boundary and $b$ a fixed boundary. Their boundary $g$-functions and the dimension of the boundary-condition-changing operator give~\cite{Cardy,CardyReview}
\begin{equation}
\frac{g_a}{g_b}=\sqrt2,\qquad \delta=\frac1{16}.
\end{equation}
Equation~\eqref{eq:main} requires $H_a\leq(1+\sqrt2)/4\simeq0.6036<1$ on the free boundary. The lowest nonvacuum primary has dimension $1/2$, consistent with the bound. Appendix~\ref{app:examples} verifies the crossing equation for this four-point function explicitly and extends this check to diagonal unitary minimal models.

The same model also illustrates the direct-sum construction. Take $a=+\oplus-$, where $+$ and $-$ denote the two fixed boundary conditions, and let $b$ be the free boundary. The spin-flip invariant operator in Eq.~\eqref{eq:sum} has dimension $\delta=1/16$, and the ratio of the boundary $g$-functions is again $g_a/g_b=\sqrt2$. Equation~\eqref{eq:main} therefore gives the same bound on $H_a$. Each fixed boundary has no relevant operator, so the bound requires a domain-wall operator between the two fixed boundaries; its exact dimension is $1/2$. The instability of the sum is thus detected even though the individual fixed boundaries are stable.

In the three-state Potts model, the same argument rules out stability of the direct sum of the three fixed boundary conditions. Let $a$ be this sum and $b$ the free boundary. Exact boundary data give~\cite{Potts}
\begin{equation}
\frac{g_a}{g_b}=\sqrt3,\qquad \delta=\frac18.
\end{equation}
Equation~\eqref{eq:main} gives $H_a\leq(3\sqrt2+\sqrt6)/8\simeq0.8365<1$. Since the individual fixed boundaries have no relevant operators, the required operator must connect two distinct fixed boundaries. Its exact dimension is $2/3$. The bound from the third derivative also excludes stability of both direct sums: stability would require $g_a/g_b\leq8/7$ for Ising and $g_a/g_b\leq24/17$ for Potts.

Table~\ref{tab:exact_stability} compares these sums with their stable tricritical counterparts~\cite{AffleckTricritical,IinoTricritical}. We restrict perturbations to preserve spin flip or cyclic permutations of the colors. Each fixed boundary has no relevant or marginal primary. The symmetry also forbids perturbations by differences of the projectors onto the fixed boundaries, which have $h=0$ and would select a particular fixed boundary. Stability is therefore determined by the lowest dimension of a domain-wall operator. The critical and tricritical examples have the same ratios of boundary $g$-functions but different dimensions $\delta$ of the boundary-condition-changing operators between fixed and free boundaries. Their contrasting stability shows why this dimension, in addition to the ratio of $g$-functions, is essential to the stability criteria.

\begin{table}[!htbp]
\centering
\small
\setlength{\tabcolsep}{6pt}
\renewcommand{\arraystretch}{1.2}
\begin{tabular}{@{}lccc@{}}
\toprule
Theory & $(\delta,g_a/g_b)$ & Exact $H_a$ & Bound on $H_a$ \\
\midrule
Ising & $(1/16,\sqrt2)$ & $1/2$: unstable & $\leq0.604$ \\
Tricritical Ising & $(7/16,\sqrt2)$ & $3/2$: stable & $\leq4.225$ \\
\addlinespace[3pt]
Three-state Potts & $(1/8,\sqrt3)$ & $2/3$: unstable & $\leq0.837$ \\
Tricritical three-state Potts & $(3/8,\sqrt3)$ & $4/3$: stable & $\leq2.510$ \\
\bottomrule
\end{tabular}
\caption{Exact checks for ordered boundaries~\cite{Cardy,CardyReview,Potts,AffleckTricritical,IinoTricritical}. In each model, $a$ is the direct sum of the fixed boundary conditions and $b$ is the free boundary. Perturbations preserve spin flip for Ising or cyclic permutations of the colors for Potts. Appendix~\ref{app:examples} verifies that $H_a$ equals the lowest dimension of a domain-wall operator. The upper bounds are from Eq.~\eqref{eq:main}.}
\label{tab:exact_stability}
\end{table}
\FloatBarrier

\section{Impurity Order in Three-dimensional \texorpdfstring{$O(N)$}{O(N)} Models}
\label{sec:o3}
The defect entropy of a single pinning field defect and the dimension of its endpoint operator constrain how many ordered impurity configurations can remain stable. We apply the bounds of Sections~\ref{sec:first}--\ref{sec:stability} to the three-dimensional $O(2)$, $O(3)$, and $O(4)$ models. We first consider orientations at equal angular intervals in an internal plane (Section~\ref{sec:oncomparison}). We then use configurations in which all pairs of orientations make a common angle, which bound the lowest dimension of a domain-wall operator over continuous ranges of relative angles (Section~\ref{sec:equalangle}), and we show that the instability criterion for a symmetric direct sum does not depend on the geometry of its orientations (Section~\ref{sec:arbitrary}). Finally, we compute the dimensions of domain-wall operators at a grid of relative angles directly on the fuzzy sphere (Section~\ref{sec:numerical}). These estimates do not use $\delta$ or $g_{\rm pin}$ and provide a consistency check of the analytic bounds, and Section~\ref{sec:physics} discusses the physical picture that emerges.

\subsection{Pinning Field Defects and Symmetry-preserving Perturbations}
\label{sec:onsetup}
A pinning field defect is obtained by coupling the bulk $O(N)$ order parameter to a magnetic field supported on a line and taking the infrared limit~\cite{AllaisSachdev,AllaisMagneticLine,ParisenToldinAssaadWessel,CuomoKomargodskiMezei,HuHeZhu,NishiokaOkuyamaShimamori,BeltonDrukkerKongStergiou}. The fixed direction $\boldsymbol n\in S^{N-1}$ of this field in the internal $O(N)$ space specifies the orientation of the defect, which we denote by $\mathcal D_{\boldsymbol n}$. It preserves the stabilizer subgroup $O(N-1)$ of $\boldsymbol n$. The broken internal rotations give $N-1$ tilt operators, which transform as a vector of $O(N-1)$ and have protected dimension one, and the broken transverse translations give the displacement operators, which transform as a transverse vector and have dimension two~\cite{CuomoKomargodskiMezei,DrukkerKongSakkas}. We take these defects to be elementary and inequivalent for distinct orientations, as in Section~\ref{sec:stability}. For $N\geq2$, internal rotations relate all such defects, so they share the same defect $g$-function $g_{\rm pin}$.

Take the trivial defect as $b$, with $g_b=1$. A defect-changing operator between $\mathcal D_{\boldsymbol n}$ and $b$ is an endpoint operator. We choose a primary endpoint operator $\psi_{\boldsymbol n}$ of dimension $\delta$ that is a scalar under transverse spatial rotations and a singlet under the full internal stabilizer $O(N-1)$ of $\boldsymbol n$, including its disconnected component. The rotated endpoints can then be chosen equivariantly, $U(R)\psi_{\boldsymbol n}U(R)^{-1}=\psi_{R\boldsymbol n}$ for $R\in O(N)$, so that every defect in the orbit carries an endpoint operator with the same $\delta$ and any finite symmetry group permuting a set of orientations permutes their endpoint operators in the same way.

Defect-changing operators between $\mathcal D_{\boldsymbol n}$ and $\mathcal D_{\boldsymbol n'}$ are domain-wall operators. Their spectrum depends only on the internal angle $\theta$ defined by $\boldsymbol n\cdot\boldsymbol n'=\cos\theta$. We denote by $\Delta_{\rm DW}(\theta)$ the lowest dimension of a scalar primary domain-wall operator. At $\theta=0$ the two defects coincide and the lowest operator is the identity. Along the family of defects generated by the tilt operators, this operator continues to a branch of domain-wall operators whose dimension vanishes quadratically as $\theta\to0$ (Section~\ref{sec:physics}).

Let $S$ be a finite set of $q$ distinct orientations permuted transitively by a finite subgroup $G\subset O(N)$, and let $a$ be the direct sum of the corresponding defects, as in Eq.~\eqref{eq:sum}. Then $g_a=qg_{\rm pin}$, the invariant combination of endpoint operators has dimension $\delta$, and its OPE with its adjoint contains only $G$-invariant operators. A primary with $0<h<1$ in this OPE gives a relevant $G$-preserving perturbation, so stability requires $H_a\geq1$. For $qg_{\rm pin}>1$, Eq.~\eqref{eq:main} gives the necessary condition
\begin{equation}
1-\frac{1}{qg_{\rm pin}}\leq2\sqrt2\,\delta.
\label{eq:onfirst}
\end{equation}
For $0<\delta\leq1/6$, the bound from the third derivative gives a stronger necessary condition by replacing the right-hand side with the first upper bound in Eq.~\eqref{eq:thirdrational}. The left-hand side increases with $q$, while the right-hand sides depend only on $\delta$. Excluding stability for a given number of orientations therefore excludes it for every larger number with the same $g_{\rm pin}$ and $\delta$. The resulting inequalities are necessary conditions for stability; they do not establish stability when satisfied, and a marginal operator with $h=1$ would require a separate analysis.

These bounds do not assume that the individual pinning field defects are stable. To identify the required relevant operator as a domain-wall operator, we additionally require the OPE on each pinning field defect to contain no relevant singlet. Any relevant primary required by crossing must then connect defects with distinct orientations, and a Hermitian combination gives a domain-wall perturbation preserving the imposed symmetry. In $O(3)$, the OPE on each pinning field defect contains only internal $SO(2)$ singlets; the dimension-one tilt operators are charged under $SO(2)$ and do not enter~\cite{CuomoKomargodskiMezei,BianchiBonomiDeSabbata}.

\subsection{Bounds on the Number of Planar Ordered Configurations}
\label{sec:oncomparison}
We first take $q$ orientations at equal angular intervals in a plane through the origin of the internal space. The rotation by $2\pi/q$ in this plane permutes the defects cyclically, and $2\pi/q$ is the angle between neighboring orientations. The direct sum describes $q$ degenerate configurations that spontaneously break the cyclic $\mathbb Z_q$ symmetry. With the reported values of $\delta$ and $g_{\rm pin}$, the bounds exclude stability for every $q\geq3$ in $O(2)$ and $O(3)$ and every $q\geq4$ in $O(4)$ (Table~\ref{tab:on_bounds}). The $O(3)$ result addresses the $120^\circ$ configuration proposed in Ref.~\cite{Yang}, whose stability is not determined by calculations for opposite orientations. The bound from the first derivative remains above one for $O(3)$ at $q=3$, so the third derivative is needed in this case. For $O(2)$ and $O(4)$, the first derivative already detects the instability. The criterion from the third derivative is the stronger test for the presence of a relevant operator; once relevance is established, the first derivative can give the lower numerical bound on $H_a$, as for $O(2)$ in Table~\ref{tab:on_bounds}.

\begin{table}[!htbp]
\centering
\small
\setlength{\tabcolsep}{8pt}
\renewcommand{\arraystretch}{1.15}
\begin{tabular}{@{}lccccc@{}}
\toprule
 & \multicolumn{2}{c}{Nominal inputs} & & \multicolumn{2}{c}{Upper bound on $H_a$} \\
\cmidrule(lr){2-3}\cmidrule(l){5-6}
Bulk CFT & $\delta$ & $g_{\rm pin}$ & $q$ & First derivative & Third derivative \\
\midrule
$O(2)$ & $0.1175$ & $0.585$ & $3$ & $\mathbf{0.77253}$ & $0.84367$ \\
$O(3)$ & $0.1330$ & $0.490$ & $3$ & $1.17657$ & $\mathbf{0.98695}$ \\
$O(4)$ & $0.1489$ & $0.439$ & $4$ & $0.97823$ & $\mathbf{0.94564}$ \\
\bottomrule
\end{tabular}
\caption{Bounds for the direct sum of $q$ pinning field defects with orientations at equal angular intervals in an internal plane. The inputs $\delta$ and $g_{\rm pin}$ are from Ref.~\cite{Yang}; their uncertainties are discussed in Section~\ref{sec:inputs}. The columns use Eq.~\eqref{eq:main} and the first upper bound in Eq.~\eqref{eq:thirdrational}, respectively, with $g_a=qg_{\rm pin}$ and $g_b=1$. The smaller bound in each row is bold. Each row gives the smallest $q$ for which stability is excluded at these inputs.}
\label{tab:on_bounds}
\end{table}

The condition for excluding three-state order in $O(3)$ will also be used to derive the bounds at other relative angles below. For $0<\delta\leq1/6$, three-state order is excluded whenever
\begin{equation}
1-\frac{1}{3g_{\rm pin}}>
\frac{\sqrt2\,\delta\left(16\delta^2-12\sqrt2\,\delta+3\sqrt2+2\right)}
{24\delta^2-6(1+2\sqrt2)\delta+1+3\sqrt2}.
\label{eq:o3region}
\end{equation}
At $(\delta,g_{\rm pin})=(0.1330,0.490)$, the necessary stability condition from Eq.~\eqref{eq:thirdrational} becomes
\begin{equation}
q\leq2.9465<3.
\label{eq:o3numbers}
\end{equation}
At $q=3$, Eq.~\eqref{eq:thirdrational} rules out $H_a\geq1$, so $\min\{H_a,H_a^3\}=H_a^3$. Taking the cube root gives
\begin{equation}
H_a\leq0.98695,\qquad q=3.
\label{eq:o3gap}
\end{equation}
The same rational criterion gives $q\leq2.305$ for $O(2)$ and $q\leq3.582$ for $O(4)$ at their nominal inputs.

The two-state cases illustrate the limitations of these bounds. Evaluating the bound from the third derivative at the central values of the Ising inputs $\delta=0.108(5)$ and $g_{\rm pin}=0.602(2)$~\cite{ZhouGaiottoHeZou,LLM} gives an upper bound of about $1.362$, which does not detect the known instability. Neither bound excludes two-state order in $O(2)$ at the inputs in Table~\ref{tab:on_bounds}. For $O(3)$ and $O(4)$, the direct sums of two pinning field defects have $2g_{\rm pin}<1$, so the condition $g_a>g_b$ needed to bound their spectra is not met. For $O(4)$ at $q=3$, the bound from the third derivative is about $1.512$, also above one.

These results extend the stability analysis beyond what is known from domain-wall operators between opposite orientations. In $O(2)$, the planar configurations contain such pairs whenever $q$ is even, and the relevant domain-wall operator between them~\cite{Yang} gives a nonzero invariant perturbation when summed over the cyclic orbit. For odd $q$, no pair has opposite orientations, so that argument does not determine stability. In $O(3)$ and $O(4)$, an irrelevant domain-wall operator between opposite orientations does not determine the dimensions of operators connecting other pairs in a multistate configuration. The ordering of defect entropies alone does not determine stability either. For $qg_{\rm pin}>1$ the trivial defect is not excluded as an infrared endpoint by the defect $g$-theorem~\cite{CuomoKomargodskiRavivMoshe}, but this inequality establishes neither the existence of such a flow nor the instability of the direct sum.

For $q\geq4$, a planar configuration contains several classes of pairs, with angles $\min(2\pi k/q,\,2\pi-2\pi k/q)$, and the bound constrains the lowest domain-wall dimension among all of them. Planar orientations at equal intervals $\theta<2\pi/q$ along an arc likewise give a direct sum covered by Eq.~\eqref{eq:weightedsum}, although not a symmetric one. In both cases, the bound constrains the minimum of $\Delta_{\rm DW}$ over the angles $\theta,2\theta,\ldots$ that occur, not $\Delta_{\rm DW}(\theta)$ itself. It becomes a bound on $\Delta_{\rm DW}(\theta)$ at the smallest separation only if $\Delta_{\rm DW}$ does not decrease with the angle. The configurations of the next subsection avoid this limitation.
\FloatBarrier

\subsection{Configurations with a Common Pairwise Angle}
\label{sec:equalangle}
The bound for a direct sum constrains $\Delta_{\rm DW}(\theta)$ at a specific angle when every pair of orientations makes the same angle $\theta$. Such configurations are obtained by tilting a regular simplex. Let $\boldsymbol e_1,\ldots,\boldsymbol e_q$ be unit vectors pointing to the vertices of a regular simplex centered at the origin of $\mathbb R^{q-1}$, so that $\boldsymbol e_i\cdot\boldsymbol e_j=-1/(q-1)$ for $i\ne j$. For $N\geq q$, embed $\mathbb R^{q-1}\times\mathbb R$ in the internal space and set
\begin{equation}
\boldsymbol n_j=\left(\sin\beta\,\boldsymbol e_j,\ \cos\beta\right),\qquad
\boldsymbol n_i\cdot\boldsymbol n_j=\cos^2\beta-\frac{\sin^2\beta}{q-1}=\cos\theta,\qquad i\ne j.
\label{eq:simplex}
\end{equation}
The symmetric group $S_q$ of the simplex acts on the first factor and permutes the orientations transitively, and all pairs are related by it. As $\beta$ ranges over $0<\beta\leq\pi/2$, the common angle covers
\begin{equation}
0<\theta\leq\theta_q:=\arccos\left(-\frac{1}{q-1}\right),
\label{eq:thetaq}
\end{equation}
with $\theta_3=120^\circ$, $\theta_4\simeq109.47^\circ$, and $\theta_5\simeq104.48^\circ$. The untilted simplex at $\beta=\pi/2$ requires only $N\geq q-1$. At fixed $N$, a continuously tilted family therefore requires $q\leq N$, whereas $q=N+1$ orientations with a common angle exist only at the regular-simplex angle $\theta_{N+1}$ and give a bound at that single angle.

For fixed $q$, the inputs to the bound are independent of $\beta$: the direct sum always contains $q$ defects, so $g_a=qg_{\rm pin}$, $g_b=1$, and $\delta$ do not change. Let $B_q$ denote the smaller of the bounds on $H_a$ from Eq.~\eqref{eq:main} and from the first upper bound in Eq.~\eqref{eq:thirdrational}. Whenever $B_q<1$, the assumption that the OPE on each pinning field defect contains no relevant singlet identifies the required operator as a domain-wall operator, and since all pairs have the same spectrum, $\Delta_{\rm DW}(\theta)\leq B_q$ for every angle realized by the configuration. When $B_q\geq1$, the bound constrains $H_a$ only and gives neither an exclusion nor a bound on $\Delta_{\rm DW}$, unless the OPE on each pinning field defect is assumed to have a gap above $B_q$. Table~\ref{tab:angular_bounds} collects the resulting bounds.

\begin{table}[!htbp]
\centering
\small
\setlength{\tabcolsep}{7pt}
\renewcommand{\arraystretch}{1.15}
\begin{tabular}{@{}lcccc@{}}
\toprule
Bulk CFT & $q$ & Range of $\theta$ & Bound & Below one for \\
\midrule
$O(2)$ & $3$ & $120^\circ$ & $\Delta_{\rm DW}\leq0.773$ & $g_{\rm pin}>0.449$ \\
\addlinespace[2pt]
$O(3)$ & $3$ & $0<\theta\leq120^\circ$ & $\Delta_{\rm DW}\leq0.987$ & $g_{\rm pin}>0.481$ \\
 & $4$ & $109.47^\circ$ & $\Delta_{\rm DW}\leq0.768$ & $g_{\rm pin}>0.361$ \\
\addlinespace[2pt]
$O(4)$ & $3$ & $0<\theta\leq120^\circ$ & $H_a\leq1.512$ (no exclusion) & --- \\
 & $4$ & $0<\theta\leq109.47^\circ$ & $\Delta_{\rm DW}\leq0.946$ & $g_{\rm pin}>0.393$ \\
 & $5$ & $104.48^\circ$ & $\Delta_{\rm DW}\leq0.774$ & $g_{\rm pin}>0.314$ \\
\bottomrule
\end{tabular}
\caption{Bounds from $q$ pinning field defects with a common pairwise angle, Eq.~\eqref{eq:simplex}, at the nominal inputs of Table~\ref{tab:on_bounds}. A range of angles requires $N\geq q$; a single angle corresponds to the untilted simplex. The bounds are the smaller of Eq.~\eqref{eq:main} and the first upper bound in Eq.~\eqref{eq:thirdrational}. Bounds below one apply to $\Delta_{\rm DW}(\theta)$ if the OPE on each pinning field defect contains no relevant singlet; for $O(4)$ at $q=3$ the bound exceeds one and constrains $H_a$ only. The last column gives the condition on $g_{\rm pin}$, at fixed nominal $\delta$, under which the bound lies below one.}
\label{tab:angular_bounds}
\end{table}

For $O(3)$ and $q=3$, the construction~\eqref{eq:simplex} reduces to three orientations at a common polar angle $\beta$ with azimuthal angles separated by $2\pi/3$:
\begin{equation}
\begin{aligned}
\boldsymbol n_j&=\left(\sin\beta\cos\frac{2\pi j}{3},\,
\sin\beta\sin\frac{2\pi j}{3},\,\cos\beta\right),\qquad j=0,1,2,\\
\boldsymbol n_i\cdot\boldsymbol n_j&=\frac{3\cos^2\beta-1}{2}=\cos\theta,
\qquad i\ne j.
\end{aligned}
\label{eq:latitude}
\end{equation}
At $\beta=\pi/2$, the three orientations lie in a plane and recover the $120^\circ$ configuration of Section~\ref{sec:oncomparison}. Decreasing $\beta$ tilts all three towards the polar axis. The angle of the rotation generating $\mathbb Z_3$ remains $2\pi/3$, whereas the actual pairwise angle $\theta$ decreases continuously towards zero. If Eq.~\eqref{eq:o3region} holds, the same assumption gives
\begin{equation}
\Delta_{\rm DW}(\theta)<1,
\qquad 0<\theta\leq\frac{2\pi}{3},
\label{eq:angleinterval}
\end{equation}
and at the nominal inputs the bound evaluates to $\Delta_{\rm DW}(\theta)\leq0.98695$ throughout this range. The bound applies without interpolation between angles or any assumption that $\Delta_{\rm DW}(\theta)$ varies monotonically with $\theta$. To test stability under the exchange symmetry of a given pair, consider a rotation by $\pi$ about the axis along the sum of their orientation vectors. This rotation exchanges the two defects, and an appropriate Hermitian combination of the relevant operators is invariant under it. The direct sum of any two pinning field defects with $0<\theta\leq120^\circ$ is therefore unstable even under perturbations preserving their exchange symmetry, although $2g_{\rm pin}<1$ prevents the bound from being applied to this pair directly. Additional microscopic symmetries could impose further selection rules on the allowed perturbations.

Four orientations pointing to the vertices of a regular tetrahedron in the internal $O(3)$ space have the common angle $\arccos(-1/3)\simeq109.47^\circ$. The direct sum has $g_a=4g_{\rm pin}$, and the bound from the first derivative gives
\begin{equation}
\Delta_{\rm DW}\!\left(\arccos(-1/3)\right)
\leq\frac{2\sqrt2\,\delta}{1-1/(4g_{\rm pin})}
\simeq0.768.
\label{eq:tetrahedron}
\end{equation}
More generally, this argument requires a relevant domain-wall operator whenever $2\sqrt2\,\delta<1-1/(4g_{\rm pin})$. At $\delta=0.133$, this requires only $g_{\rm pin}>0.40076$, about $18\%$ below the nominal value; the bound from the third derivative relaxes the condition further to $g_{\rm pin}>0.361$. Since $q=4=N+1$ in $O(3)$, this configuration cannot be tilted within the internal space, and the bound holds only at the tetrahedral angle.

In $O(4)$, three orientations with a common angle give no exclusion, but four do. The tilted tetrahedron~\eqref{eq:simplex} with $q=4$ fits into the four-dimensional internal space and gives
\begin{equation}
\Delta_{\rm DW}(\theta)\leq0.94564<1,\qquad 0<\theta\leq\arccos(-1/3),
\label{eq:o4interval}
\end{equation}
from the bound from the third derivative. This exclusion holds for $g_{\rm pin}>0.39312$ at fixed $\delta=0.1489$ or for $\delta<0.16518$ at fixed $g_{\rm pin}=0.439$, margins of about $10\%$ in each direction. The five vertices of a regular four-simplex give the stronger bound $\Delta_{\rm DW}(104.48^\circ)\leq0.774$ at that single angle.

\subsection{Configurations of Arbitrary Shape}
\label{sec:arbitrary}
The instability criterion for a symmetric direct sum does not depend on the geometry of its orientations. For any finite set of $q$ distinct orientations permuted transitively by a finite group $G\subset O(N)$, with the equivariant endpoint operators of Section~\ref{sec:onsetup}, the invariant endpoint operator has dimension $\delta$ and the direct sum has $g_a=qg_{\rm pin}$, whatever the angles between the orientations. The crossing inequality therefore requires a relevant $G$-invariant primary whenever $1-1/(qg_{\rm pin})$ exceeds the first upper bound in Eq.~\eqref{eq:thirdrational}, and an invariant Hermitian component of this primary is a relevant $G$-preserving perturbation. Under the assumption that the OPE on each pinning field defect contains no relevant singlet, this primary is a domain-wall operator. At the nominal inputs, this excludes stable impurity order for every symmetric configuration with $q\geq3$ orientations in $O(2)$ and $O(3)$ and with $q\geq4$ in $O(4)$, including the octahedral, cubic, and icosahedral configurations in $O(3)$, without any planar restriction. Compared with Section~\ref{sec:oncomparison}, the number of configurations excluded is unchanged, but no restriction on their geometry remains.

The equal-angle bounds of Section~\ref{sec:equalangle} give a separate, pairwise statement. For unit vectors $\boldsymbol n_1,\ldots,\boldsymbol n_q$,
\begin{equation}
0\leq\Bigl|\sum_{i=1}^q\boldsymbol n_i\Bigr|^2=q+2\sum_{i<j}\boldsymbol n_i\cdot\boldsymbol n_j,
\end{equation}
so the average of $\boldsymbol n_i\cdot\boldsymbol n_j$ over pairs is at least $-1/(q-1)$, and some pair makes an angle of at most $\theta_q$ in Eq.~\eqref{eq:thetaq}. Since $\theta_q\leq\theta_3=120^\circ$ for all $q\geq3$, every configuration of three or more distinct orientations contains a pair at an angle of at most $120^\circ$, and every configuration of four or more contains a pair at an angle of at most $109.47^\circ$. The bound on the smallest pair angle is saturated by a regular simplex when $q\leq N+1$. Combining this inequality with Eqs.~\eqref{eq:angleinterval} and \eqref{eq:o4interval} gives a relevant scalar domain-wall operator between some pair of any configuration of $q\geq3$ orientations in $O(3)$ and of $q\geq4$ orientations in $O(4)$. This spectral statement about a pair and the instability of the symmetric direct sum derived above are logically distinct: the former locates a relevant operator in a specific pair sector, while the latter guarantees a relevant invariant perturbation of the configuration as a whole.

\subsection{Numerical Inputs and Uncertainties}
\label{sec:inputs}
Tables~\ref{tab:on_bounds} and \ref{tab:angular_bounds} combine estimates of the dimensions of endpoint operators from finite systems with extrapolated defect $g$-functions. The values of $\delta$ are the estimates for the largest systems studied in Tables 3--5 of Ref.~\cite{Yang}: $0.1175$ at 11 orbitals for $O(2)$, $0.1330$ at 10 orbitals for $O(3)$, and $0.1489$ at 9 orbitals for $O(4)$. The corresponding $g_{\rm pin}=0.585,0.490,0.439$ are the extrapolated estimates in their Table 2, for which no error intervals are quoted. The estimates of $\delta$ show weak size dependence, for example $0.1330$ for 8, 9, or 10 orbitals and $0.1333$ for 7 orbitals in $O(3)$~\cite{Yang}. The lowest nonvacuum singlet primaries on a pinning field defect are reported to have dimensions $1.609(11)$, $1.635(58)$, and $1.666(38)$ for $N=2,3,4$, supporting the absence of relevant singlets assumed in Section~\ref{sec:onsetup}~\cite{Yang}. The endpoint data are assumed to correspond to a singlet of the full $O(N-1)$ stabilizer, including its disconnected component, as required in Section~\ref{sec:onsetup}.

All exclusions quoted from these inputs are conditional on the continuum values of $\delta$ and $g_{\rm pin}$ satisfying the stated inequalities. Establishing an exclusion in the continuum requires verifying the relevant inequality throughout a joint uncertainty region in $(\delta,g_{\rm pin})$. If this region is a rectangle contained in $0<\delta\leq1/6$ and $qg_{\rm pin}>1$, it suffices to test its least favorable corner, the largest $\delta$ and smallest $g_{\rm pin}$: the left-hand side $1-1/(qg_{\rm pin})$ of the criteria in Section~\ref{sec:onsetup} increases with $g_{\rm pin}$, while both right-hand sides increase with $\delta$, as proved for the bound from the third derivative in Appendix~\ref{app:third}. The one-parameter margins quoted below are sensitivity diagnostics, not joint confidence intervals, and they differ considerably between the applications. In $O(3)$, Eq.~\eqref{eq:o3region} holds for $\delta<0.13664$ at fixed $g_{\rm pin}=0.490$, or for $g_{\rm pin}>0.48126$ at fixed $\delta=0.133$, margins of $2.7\%$ and $1.8\%$. The tetrahedral bound in $O(3)$, the tilted tetrahedron in $O(4)$, and the triangle in $O(2)$ tolerate deviations of $10$--$26\%$ in $g_{\rm pin}$ (Table~\ref{tab:angular_bounds}). The three-state exclusion in $O(3)$ is thus the most sensitive to the input data. The direct computation below provides a check of it that does not use these inputs.

\subsection{Direct Computation of Domain-wall Dimensions on the Fuzzy Sphere}
\label{sec:numerical}
We now estimate $\Delta_{\rm DW}(\theta)$ directly on the angular grid $\theta=0^\circ,15^\circ,\ldots,180^\circ$ in steps of $15^\circ$, supplemented by $72^\circ$ and the tetrahedral angle $109.47^\circ$. The energy estimator defined below does not use $\delta$ or $g_{\rm pin}$, although it shares the regulator and the bulk tuning with the calculations that provide those inputs.

\paragraph{Model and pinning field defects.}
We use exact diagonalization in the fuzzy-sphere regularization~\cite{ZhuFuzzySphere} of the $O(N)$ Wilson--Fisher CFT constructed in Ref.~\cite{GuoZhouWeiHe}. Fermions with $N+1$ flavors $\psi_\alpha$, $\alpha=0,\ldots,N$, occupy the lowest Landau level of a monopole of charge $s$, which has $N_{\rm orb}=2s+1$ orbitals labeled by the angular momentum $m=-s,\ldots,s$ about the axis through the poles, with total particle number $N_e=N_{\rm orb}$. The Hamiltonian is
\begin{equation}
H=\int d^2x\left[n_{\rm tot}^2-U_1\sum_{a=1}^N V_a\nabla^2V_a+h_1n_{00}\right],\qquad
V_a=\psi_a^\dagger\psi_0+\psi_0^\dagger\psi_a,
\label{eq:fuzzyH}
\end{equation}
where $V_a$ is the $O(N)$ order parameter, $n_{00}=2\psi_0^\dagger\psi_0$, and $n_{\rm tot}=\sum_\alpha\psi_\alpha^\dagger\psi_\alpha$. We use the field normalization and the orbital form of the projected interaction of Ref.~\cite{GuoZhouWeiHe}, with the nominal couplings $(U_1,h_1)=(0.270,-0.084)$, $(0.264,-0.096)$, and $(0.255,-0.106)$ for $N=2,3,4$, which were optimized there at the largest accessible system sizes by minimizing the deviation of the spectrum from the expected conformal tower. This is the bulk model used in Ref.~\cite{Yang}. Under radial quantization, a straight line defect through the origin maps to a pair of point defects at the poles of the sphere~\cite{HuHeZhu}. In the lowest Landau level, a field localized at a pole acts only on the orbital with $m=\pm s$, so we realize the two pinning field defects by
\begin{equation}
H_d=h_d\sum_{a=1}^N\left[n_1^a\bigl(c_{s,a}^\dagger c_{s,0}+\text{h.c.}\bigr)+n_2^a\bigl(c_{-s,a}^\dagger c_{-s,0}+\text{h.c.}\bigr)\right],
\label{eq:pinning}
\end{equation}
where $c_{m,\alpha}$ annihilates a fermion of flavor $\alpha$ in orbital $m$, $\boldsymbol n_1$ and $\boldsymbol n_2$ are the orientations at the north and south poles, and $h_d=1000$ as in Ref.~\cite{Yang}. We checked that the estimates below change by less than $0.2\%$ when $h_d$ is reduced to $20$. For $\boldsymbol n_1=\boldsymbol n_2$ the low-lying states are local operators on a single pinning field defect, while for $\boldsymbol n_1\ne\boldsymbol n_2$ they are domain-wall operators.

\paragraph{Extraction of dimensions and symmetries.}
We choose $\boldsymbol n_{1,2}=(\cos\tfrac\theta2,\pm\sin\tfrac\theta2,0,\ldots)$ and define the finite-size estimator
\begin{equation}
\Delta_{\rm DW}^{(N_{\rm orb})}(\theta)=3\,\frac{E_0^{\boldsymbol n_1\boldsymbol n_2}-E_0^{\boldsymbol n_1\boldsymbol n_1}}{E_T-E_0},
\label{eq:dwextraction}
\end{equation}
where $E_0^{\boldsymbol n_1\boldsymbol n_2}$ is the lowest energy with the two pinning fields in the sector with vanishing angular momentum $L_z$ about the axis through the poles, which contains the scalar domain-wall primaries, and $E_0$ and $E_T$ are the energies of the vacuum and the stress tensor without defects, which fix the normalization through $\Delta_T=3$. At $\theta=\pi$ this is the estimator of Ref.~\cite{Yang}. Its continuum limit gives the lowest scalar domain-wall dimension, provided the bulk and the pole defects flow to the intended fixed points. For every $\theta$, the Hamiltonian is invariant under a $\mathbb Z_2$ that combines the spatial rotation by $\pi$ exchanging the poles with the internal reflection exchanging $\boldsymbol n_1$ and $\boldsymbol n_2$; we diagonalize both eigenvalue sectors of this symmetry within $L_z=0$ and take the lower energy. At a generic angle the residual internal symmetry is the $O(N-2)$ acting on the directions orthogonal to the plane of $\boldsymbol n_1$ and $\boldsymbol n_2$. The ground states lie in the sector even under the reflection of each orthogonal direction. For $N=4$ we verified that the lowest states of the other reflection-parity sectors are separated from this ground state by a finite gap, so it carries zero charge under the residual $SO(2)$ and is a singlet of the residual $O(2)$; for $N=3$ the residual symmetry is the reflection itself. The defect-free energies $E_0$ and $E_T$ are computed using the $U(1)$ subgroup of $O(N)$ acting on $(\psi_1,\psi_2)$ and the internal reflections. $E_T$ is the lowest energy in the sector with $L_z=2$, vanishing $U(1)$ charge, and even parity under the improper reflection and under the reflections of the components $a\geq3$; the competing states $\partial J$ and $\partial\partial\phi_a$ are odd under one of these reflections, so the stress tensor is the lowest state of this sector, and the resulting normalization reproduces the bulk vector dimension $\Delta_\phi\simeq0.516$, $0.516$, and $0.514$ at the largest sizes for $N=2,3,4$, compared with $0.519$, $0.519$, and $0.518$ from the conformal bootstrap.

\paragraph{Comparison with earlier results at $\theta=\pi$.}
At common system sizes, our finite-size estimates of $\Delta_{\rm DW}(\pi)$ differ from those of Ref.~\cite{Yang} by $+0.2$--$0.3\%$ for $O(3)$ and $O(4)$ and by $-1.3$--$1.7\%$ for $O(2)$; for example, we obtain $0.9066$ against $0.9041$ for $O(3)$ at $N_{\rm orb}=10$. Both calculations use the estimator~\eqref{eq:dwextraction} with the same nominal couplings, and the differences are insensitive to $h_d$. We attribute them to differences in the implementation of the projected interaction, to which finite-size estimates are sensitive through the location of the bulk critical point; a comparison of the four energies entering Eq.~\eqref{eq:dwextraction} under identical conditions would be needed to resolve them. They are far smaller than the margins relevant below and do not affect any conclusion.

\begin{figure}[!t]
\centering
\includegraphics[width=\textwidth]{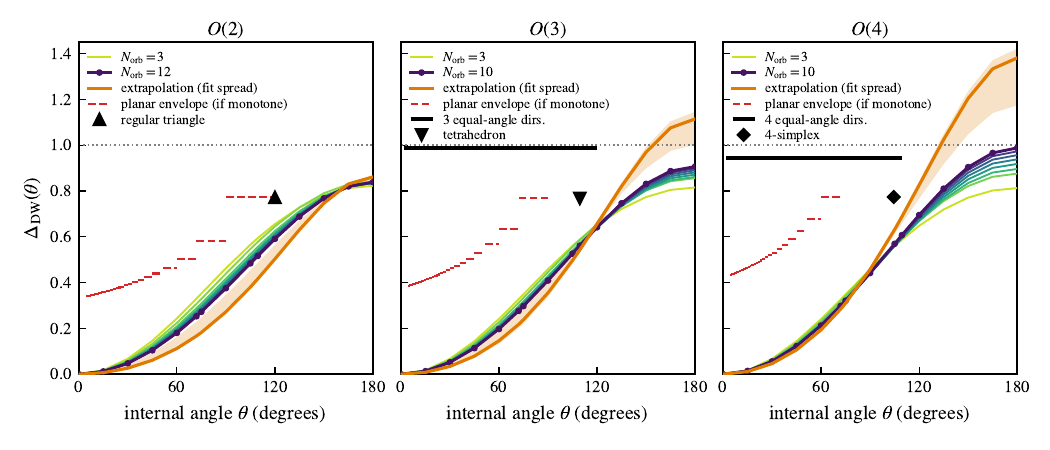}
\caption{Finite-size estimates of the lowest dimension of a scalar domain-wall operator between two pinning field defects with relative internal angle $\theta$, computed on the fuzzy sphere for $N_{\rm orb}=3,\ldots,12$ ($O(2)$) and $3,\ldots,10$ ($O(3)$, $O(4)$); darker curves correspond to larger systems, and lines connect the sampled angles. Orange curves are the continuum estimates $\Delta_\infty(\theta)$ from fits $\Delta_\infty(\theta)+b(\theta)N_{\rm orb}^{-\omega/2}$ to the four largest sizes, and orange bands show the spread of the fits described in Table~\ref{tab:dwnumerics}. Black lines and symbols are the equal-angle bounds of Table~\ref{tab:angular_bounds} that lie below one; they apply to $\Delta_{\rm DW}(\theta)$ on the stated intervals, with $\theta=0$ excluded, and at the isolated simplex angles. The red dashed staircase shows the bounds from planar configurations with $q$ orientations at equal intervals $\theta\leq2\pi/q$; these bound the minimum of $\Delta_{\rm DW}$ over the angles occurring in the configuration and bound $\Delta_{\rm DW}(\theta)$ itself only if $\Delta_{\rm DW}$ does not decrease with the angle. All bounds use the nominal inputs of Table~\ref{tab:on_bounds}. The dotted line marks the threshold $\Delta_{\rm DW}=1$.}
\label{fig:dwnumerics}
\end{figure}

\paragraph{Results.}
Figure~\ref{fig:dwnumerics} shows the estimates for all system sizes. Four features are common to all three models. First, $\Delta_{\rm DW}^{(N_{\rm orb})}(\theta)$ increases monotonically with $\theta$ on the sampled grid at every system size. Second, it is consistent with quadratic behavior at small angles: the ratio $\Delta_{\rm DW}(15^\circ)/\Delta_{\rm DW}(30^\circ)$ equals $0.253$ at the largest sizes, close to $1/4$. Third, at the sampled angles within the stated domains, the finite-size estimates lie below the equal-angle bounds, and they also lie below the planar envelope interpreted as a pointwise bound. Fourth, the finite-size dependence is concentrated at large angles. For $\theta\leq120^\circ$, the values at the two largest sizes differ by about $1\%$ or less, while at $\theta=180^\circ$ they drift upwards with $N_{\rm orb}$ in $O(3)$ and $O(4)$, as found in Ref.~\cite{Yang}.

\begin{table}[!t]
\centering
\small
\setlength{\tabcolsep}{4.2pt}
\renewcommand{\arraystretch}{1.15}
\begin{tabular}{@{}lcccccc@{}}
\toprule
 & \multicolumn{2}{c}{$O(2)$} & \multicolumn{2}{c}{$O(3)$} & \multicolumn{2}{c}{$O(4)$} \\
\cmidrule(lr){2-3}\cmidrule(lr){4-5}\cmidrule(l){6-7}
$\theta$ & $N_{\rm orb}=12$ & Fit spread & $N_{\rm orb}=10$ & Fit spread & $N_{\rm orb}=10$ & Fit spread \\
\midrule
$60^\circ$ & 0.180 & 0.11--0.16 & 0.197 & 0.13--0.19 & 0.213 & 0.19--0.22 \\
$72^\circ$ & 0.253 & 0.16--0.23 & 0.276 & 0.20--0.27 & 0.298 & 0.27--0.31 \\
$90^\circ$ & 0.376 & 0.27--0.35 & 0.409 & 0.34--0.41 & 0.442 & 0.45--0.47 \\
$105^\circ$ & 0.484 & 0.37--0.45 & 0.526 & 0.48--0.53 & 0.569 & 0.61--0.64 \\
$109.47^\circ$ & 0.516 & 0.41--0.49 & 0.561 & 0.53--0.57 & 0.607 & 0.66--0.70 \\
$120^\circ$ & 0.591 & 0.50--0.56 & 0.642 & 0.65--0.67 & 0.695 & 0.77--0.84 \\
$150^\circ$ & 0.768 & 0.75--0.76 & 0.831 & 0.90--0.99 & 0.904 & 1.05--1.24 \\
$180^\circ$ & 0.839 & 0.85--0.86 & 0.907 & 1.00--1.15 & 0.988 & 1.17--1.42 \\
\bottomrule
\end{tabular}
\caption{Finite-size estimates of the lowest scalar domain-wall dimension at the largest system sizes and the spread of the continuum fits. The baseline fit at each angle is $\Delta_\infty(\theta)+b(\theta)N_{\rm orb}^{-\omega/2}$, where $\omega=\Delta_{\hat\phi_1}-1=0.609(11)$, $0.635(58)$, $0.666(38)$ for $N=2,3,4$, with $\hat\phi_1$ the lowest irrelevant internal-singlet primary on the pinning field defect~\cite{Yang}. The spread covers this fit for the central and extreme values of $\omega$ applied to the largest three, four, and five sizes, and quadratic fits in $1/N_{\rm orb}$ to the largest four and five sizes. The ranges are fit-variation envelopes, not statistical confidence intervals or rigorous bounds on the continuum limit.}
\label{tab:dwnumerics}
\end{table}

\paragraph{Extrapolation.}
Our baseline continuum fit assumes that the leading finite-size correction is generated by the lowest irrelevant internal-singlet primary $\hat\phi_1$ on the pinning field defect, $\Delta^{(N_{\rm orb})}_{\rm DW}(\theta)=\Delta_\infty(\theta)+b(\theta)\,N_{\rm orb}^{-\omega/2}$ with $\omega=\Delta_{\hat\phi_1}-1$ and fit parameters $\Delta_\infty(\theta)$ and $b(\theta)$ at each angle, following Ref.~\cite{Yang}, which includes a second irrelevant correction that we omit. We vary $\omega$ within its quoted uncertainty, the number of sizes included, and the form of the fit. Table~\ref{tab:dwnumerics} gives the resulting spread, and Fig.~\ref{fig:dwnumerics} shows the fitted curves. For the sampled angles up to $120^\circ$, the finite-size estimates and all fits considered support a relevant scalar domain-wall operator: $\Delta_{\rm DW}(120^\circ)$ lies in $0.50$--$0.56$ for $O(2)$, $0.65$--$0.67$ for $O(3)$, and $0.77$--$0.84$ for $O(4)$. Small changes between neighboring sizes alone do not bound the remaining continuum correction; with a correction exponent $\omega/2\simeq0.3$, a difference of $1\%$ between $N_{\rm orb}=9$ and $10$ corresponds to a remaining shift of several tens of percent, which is the origin of the spread in the table. At large angles the extrapolation is more sensitive. At $180^\circ$, the fits give $0.85$--$0.86$ for $O(2)$, $1.00$--$1.15$ for $O(3)$, and $1.17$--$1.42$ for $O(4)$, in line with Ref.~\cite{Yang}; the $O(3)$ range touches the relevance threshold, so strict irrelevance of the wall between opposite orientations is not resolved by these fits. Every fitted curve for $O(3)$ and $O(4)$ crosses one, at angles $\theta_*\simeq151^\circ$--$179^\circ$ and $\theta_*\simeq131^\circ$--$144^\circ$, respectively, obtained by linear interpolation between sampled angles; no fitted curve for $O(2)$ crosses one. These ranges describe the variation among the fits, not confidence intervals.

\paragraph{Comparison with the analytic bounds.}
The direct estimates are consistent with each exclusion in Tables~\ref{tab:on_bounds} and \ref{tab:angular_bounds}. In $O(3)$, $\Delta^{(10)}_{\rm DW}(120^\circ)=0.642$ and the fitted range $0.65$--$0.67$ lie about $35\%$ and $32$--$34\%$ below the bound $0.987$, and $\Delta^{(10)}_{\rm DW}(109.47^\circ)=0.561$, with fitted range $0.53$--$0.57$, lies below the tetrahedral bound $\simeq0.768$. In $O(2)$, $\Delta^{(12)}_{\rm DW}(120^\circ)=0.591$ and the fitted range $0.50$--$0.56$ lie below the bound $0.773$ from the triangle. In $O(4)$, $\Delta^{(10)}_{\rm DW}(\theta)\leq0.607$ at the sampled angles up to $109.47^\circ$, with fitted ranges up to $0.70$, compared with the bound $0.946$; the value $\Delta^{(10)}_{\rm DW}(105^\circ)=0.569$ is close to, but does not directly test, the isolated four-simplex bound $0.774$ at $\arccos(-1/4)\simeq104.48^\circ$. A finite-size estimate below a bound is a consistency check, and the fitted ranges indicate that the same holds for the continuum values. The three-state exclusion in $O(3)$ is the most sensitive to the input data, with a margin of $2\%$ in $g_{\rm pin}$ (Section~\ref{sec:inputs}); the direct estimates at $120^\circ$ lie more than $30\%$ below the relevance threshold at all sizes and in all fits, so this conclusion does not hinge on the precise values of $\delta$ and $g_{\rm pin}$. Subject to the extrapolation and symmetry-selection assumptions, the direct estimates also support instability of three-state order in $O(4)$, where $\Delta_{\rm DW}(120^\circ)\leq0.84$ in all fits; this case lies beyond the reach of the analytic bound.

\subsection{Physical Implications}
\label{sec:physics}
\paragraph{Monotonicity of the bounds and of the dimensions.}
The two families of bounds behave differently as functions of the angle. For $q\leq N$, tilting the simplex in Eq.~\eqref{eq:simplex} extends the same bound $B_q$ to every angle in $0<\theta\leq\theta_q$; it does not change the bound at fixed $q$. At fixed $N$, the number of orientations with a common angle cannot exceed $N+1$, and the configuration with $q=N+1$ gives a bound only at the regular-simplex angle. The equal-angle bounds therefore do not form a monotone pointwise envelope: in $O(3)$, the bound $0.768$ at the tetrahedral angle is stronger than the bound $0.987$ available at slightly smaller angles. The planar construction behaves differently. A planar configuration of $q$ orientations has neighboring angles of at most $2\pi/q$, so more orientations fit at smaller spacings, and the corresponding bound decreases towards $2\sqrt2\,\delta$ as the spacing tends to zero. This staircase, shown in Fig.~\ref{fig:dwnumerics}, bounds the minimum of $\Delta_{\rm DW}$ over the angles present in the configuration. It becomes a pointwise bound on $\Delta_{\rm DW}(\theta)$ at the smallest spacing only under the additional assumption that $\Delta_{\rm DW}$ does not decrease with the angle. The equal-angle bounds and the instability criterion of Section~\ref{sec:arbitrary} do not rely on this assumption.

One might expect such monotonicity to be automatic. A tempting argument runs as follows: rotating the orientation continuously from $\boldsymbol n$ to $\boldsymbol n'$ passes through every intermediate orientation, so a domain wall at angle $\theta$ might be expected to be at least as costly as one at any smaller angle. Our $O(N)$ data are indeed monotonic on the sampled grid. The argument is nevertheless not a general principle, as the $\mathbb Z_k$ parafermion CFTs in two dimensions show~\cite{FateevZamolodchikov}. Their A-type Cardy boundary states $(l,m)$~\cite{MaldacenaMooreSeiberg} include fixed boundary conditions $(0,2j)$, in which the $\mathbb Z_k$ spin is fixed to the orientation $2\pi j/k$, and the boundary conditions $(1,2j+1)$ associated with chords between adjacent orientations; for $k=3$ these are the fixed and mixed boundary conditions of the three-state Potts model~\cite{Cardy,Potts}. Each boundary condition can be assigned the orientation $\pi m/k$ on the $\mathbb Z_k$ circle. The boundary-condition-changing operator from $(0,0)$ to $(l,m)$ has the dimension of the parafermion primary $\Phi^l_m$. Table~\ref{tab:parafermion} lists these dimensions as functions of the relative orientation. They are not monotonic. In the three-state Potts model, the operator at $120^\circ$ has dimension $2/3$, while the one at $180^\circ$, between the fixed boundary $A$ and the mixed boundary $BC$, has dimension $2/5$. The $\mathbb Z_4$ parafermion theory, a $c=1$ free-boson orbifold~\cite{DijkgraafVerlindeVerlinde}, has $3/4$ at $90^\circ$ but $9/16$ at $135^\circ$.

\begin{table}[!htbp]
\centering
\small
\setlength{\tabcolsep}{6pt}
\renewcommand{\arraystretch}{1.18}
\begin{tabular}{@{}lcccccc@{}}
\toprule
Theory & \multicolumn{6}{c}{Relative orientation and dimension of the boundary-condition-changing operator} \\
\midrule
$\mathbb Z_3$ (three-state Potts) & $60^\circ$: $\tfrac1{15}$ & $120^\circ$: $\tfrac23$ & $180^\circ$: $\tfrac25$ & & & \\
$\mathbb Z_4$ ($c=1$ orbifold) & $45^\circ$: $\tfrac1{16}$ & $90^\circ$: $\tfrac34$ & $135^\circ$: $\tfrac9{16}$ & $180^\circ$: $1$ & & \\
$\mathbb Z_5$ & $36^\circ$: $\tfrac2{35}$ & $72^\circ$: $\tfrac45$ & $108^\circ$: $\tfrac{23}{35}$ & $144^\circ$: $\tfrac65$ & $180^\circ$: $\tfrac67$ & \\
\bottomrule
\end{tabular}
\caption{Dimensions of boundary-condition-changing operators between the fixed boundary at orientation $0$ and the boundary conditions $(0,2j)$ (even multiples of $\pi/k$) and $(1,2j+1)$ (odd multiples) of the $\mathbb Z_k$ parafermion CFTs. They equal $h^{l'}_{m'}=l'(l'+2)/[4(k+2)]-m'^2/(4k)$ evaluated on the representative $(l',m')$ with $|m'|\leq l'$ of the field-identification class $(l,m)\sim(k-l,m+k)$, and are not monotonic in the relative orientation.}
\label{tab:parafermion}
\end{table}

These examples also show why intermediate orientations do not control the dimension. Fusing two domain walls between neighboring orientations does not produce a wall whose dimension is bounded by the sum: in the Potts model, two walls of dimension $1/15$ fuse into the wall between $A$ and $B$, of dimension $2/3>2/15$. Assigning angular labels to boundary conditions therefore does not by itself imply a triangle inequality or monotonicity. The boundary conditions in these examples are not related by a continuous symmetry and do not share a $g$-function, unlike the pinning field defects of the $O(N)$ models, so they do not disprove a monotonicity property restricted to the $O(N)$ orbit. Whether $\Delta_{\rm DW}(\theta)$ is monotonic there is a dynamical question; our finite-size data support it on the sampled grid for $N=2,3,4$, but it does not follow from kinematics. The equal-angle bounds and the direct-sum criterion do not assume it, whereas the pointwise interpretation of the planar envelope does.

\paragraph{Information provided by the bounds.}
Several features show that the bounds contain information not available from simpler considerations.
\begin{itemize}
\item They use only the defect entropy of a single pinning field defect and the dimension of its endpoint operator. No domain-wall dimension enters at any angle, and the same two numbers constrain all symmetric configurations with a given number of orientations at once.
\item They cannot be inferred from the domain wall between opposite orientations. In $O(3)$ and $O(4)$ that wall is irrelevant according to the extrapolations of Ref.~\cite{Yang}, so even a monotonic $\Delta_{\rm DW}(\theta)$ would leave its value at $120^\circ$ undetermined. In $O(2)$ the relevant opposite-orientation wall implies the instability of odd-$q$ order only if monotonicity is assumed, and the examples above show that monotonicity is not automatic. The bounds settle these cases directly.
\item They distinguish fixed points with the same ratio of $g$-functions. The critical and tricritical examples in Table~\ref{tab:exact_stability} have equal ratios but opposite stability.
\item The ordering of defect entropies is not enough on its own. The criterion $2g_{\rm pin}<1$ discussed in Ref.~\cite{Yang} excludes the trivial defect as an infrared endpoint of two-state order for $N\geq3$, and $qg_{\rm pin}>1$ leaves it allowed for $q\geq3$, but neither implies stability or instability. The bounds convert the excess $qg_{\rm pin}-1$, together with $\delta$, into the existence of a relevant operator.
\item The direct estimates are consistent with the bounds wherever a bound lies below one, at all accessible system sizes. Lying below an upper bound is a consistency check rather than a proof of the bound, but a violation at these margins would have signaled a failure of one of the assumptions.
\end{itemize}

\paragraph{Critical angle and the number of ordered configurations.}
The direct estimates suggest a simple picture of impurity order built from the pinning field defects considered here. Suppose that the lowest symmetry-allowed domain-wall branch is monotonic in the angle and crosses one once, at an angle $\theta_*$. Strict irrelevance of all domain-wall perturbations then requires every pair of orientations to be separated by more than $\theta_*$, with equality a marginal case requiring a separate analysis. Since every configuration of three or more orientations contains a pair at an angle of at most $120^\circ$ (Section~\ref{sec:arbitrary}), strictly stable order with $q\geq3$ requires $\theta_*<120^\circ$. Our fits give no crossing for $O(2)$, $\theta_*\simeq151^\circ$--$179^\circ$ for $O(3)$, and $\theta_*\simeq131^\circ$--$144^\circ$ for $O(4)$. Under the stated symmetry and extrapolation assumptions, the data therefore support the exclusion of these direct-sum candidates with $q\geq2$ in $O(2)$ and $q\geq3$ in $O(3)$ and $O(4)$. Two-state configurations at sufficiently large separation remain candidates, of which antipodal easy-axis order is a particularly symmetric example; in $O(3)$ its domain wall is close to the relevance threshold. The analytic bounds establish the exclusions for $q\geq3$ in $O(2)$ and $O(3)$ and for $q\geq4$ in $O(4)$ using only the inputs of Table~\ref{tab:on_bounds}; the direct estimates add $q=3$ in $O(4)$ and $q=2$ in $O(2)$ as numerical support. These statements concern finite direct sums of pinning field defects and do not classify all possible impurity fixed points. Since $\theta_*$ decreases with $N$, three-state order may become possible at larger $N$, once $\Delta_{\rm DW}(120^\circ)$ exceeds one. A linear continuation of the three available estimates, $0.50$--$0.56$, $0.65$--$0.67$, and $0.77$--$0.84$, would place this around $N\approx5$ or $6$, but this is only a heuristic; it motivates direct calculations at $N=5$ and $6$ and does not by itself establish a stable three-state fixed point, which would also require the absence of other relevant perturbations. The analytic bound on the number of planar configurations, $q\leq2.31$, $2.95$, and $3.58$ for $N=2,3,4$, grows with $N$ in the same direction.

\paragraph{Dependence on $N$: rigidity of the pinned orientation.}
At every sampled angle, the estimated dimensions increase with $N$. At $120^\circ$ and the largest sizes, for example, $\Delta_{\rm DW}=0.591$, $0.642$, and $0.695$ for $N=2,3,4$. This trend has a natural interpretation. Two conjugate domain walls separated by a distance $\tau$ along the line carry a weight $\tau^{-2\Delta_{\rm DW}}$, which corresponds to a logarithmic separation cost $2\Delta_{\rm DW}\log(\tau/\epsilon)$, with $\epsilon$ a short-distance cutoff along the defect, and to the renormalization-group eigenvalue $1-\Delta_{\rm DW}$ of the corresponding coupling in Section~\ref{sec:stability}. At the level of a single pair, this is the mechanism by which long-range interactions stabilize order in one-dimensional Ising chains with $1/r^2$ couplings~\cite{Thouless,AndersonYuvalHamann,Cardy1981}; a full many-wall description would require higher-point defect correlators. Here the logarithmic interaction is mediated by the critical bulk, and its strength measures how rigidly the bulk holds the pinned orientation. At large $N$, the pinning field defect is described by a saddle point on which the order parameter is of order $\sqrt N$~\cite{CuomoKomargodskiMezei}. If the junction between two orientations has a corresponding saddle with positive excess action, the dimensions of domain-wall operators grow as $\Delta_{\rm DW}(\theta)=Nf(\theta)+O(1)$ at fixed nonzero angle, in line with the linear growth in $N$ of the $4-\epsilon$ result of Ref.~\cite{CuomoHeKomargodski}; a controlled calculation of $f(\theta)$ in three dimensions would be needed to establish the angular dependence. In this regime a domain wall at fixed angle becomes irrelevant and $\theta_*$ decreases with $N$. At small angles the same rigidity appears in conformal perturbation theory. With the angular coupling normalized as $\delta S=\theta\int d\tau\,t(\tau)$ and $\langle t(\tau)t(0)\rangle=C_t/\tau^2$, a rotation by a small angle over a segment of length $L$ changes the free energy by $C_t\theta^2\log(L/\epsilon)$, and comparing with the two junctions at its ends gives
\begin{equation}
\Delta_{\rm DW}(\theta)=\frac{C_t}{2}\,\theta^2+O(\theta^4),
\label{eq:smallangle}
\end{equation}
with $\theta$ in radians~\cite{CuomoHeKomargodski}. The ratios $\Delta_{\rm DW}(\theta)/\theta^2$ at $\theta=15^\circ$ and the largest sizes, $0.174$, $0.191$, and $0.205$ for $N=2,3,4$, are finite-size, finite-angle estimates of $C_t/2$ and show the same ordering in $N$. The defect data of Ref.~\cite{Yang} follow the same trend: $\delta$ increases and $g_{\rm pin}$ decreases with $N$, which increases the bound $q_{\max}=1/[g_{\rm pin}(1-R(\delta))]$ on the number of planar configurations, where $R(\delta)$ denotes the first upper bound in Eq.~\eqref{eq:thirdrational}. Interpreting the growth of the domain-wall cost with $N$ as increasing semiclassical rigidity is qualitative at these small values of $N$; the measured ordering is the controlled statement.

\paragraph{Infrared endpoints.}
The bounds establish instability but do not determine where the resulting flows end. The defect $g$-theorem~\cite{CuomoKomargodskiRavivMoshe} requires the infrared defect to have a smaller $g$-function. For regular planar configurations, whose orientations sum to zero, the trivial defect is compatible with this inequality for $q\geq3$. The allowed and stable endpoints depend on the full microscopic symmetry, including its action on the directions orthogonal to the plane: cyclic symmetry alone leaves a magnetic field along such a direction invariant, so a pinning field defect orthogonal to the plane is a candidate as well. For the tilted configurations in Eq.~\eqref{eq:simplex} with $\beta<\pi/2$, the orientations have a nonzero average along the axis of the simplex, and a single pinning field defect along that axis, with $g_{\rm pin}<qg_{\rm pin}$, is a symmetry-compatible candidate with smaller $g$. The domain-wall perturbation may then lock the impurity to the average orientation, but the renormalization-group trajectory, including the sign of the selected axial orientation, is not fixed by crossing or by $g$-monotonicity. For two-state order in $O(3)$ and $O(4)$, $2g_{\rm pin}<1$ rules out the trivial endpoint but does not by itself establish stability. Following these flows numerically, for example by adding to a direct sum realized by an impurity spin coupled to the bulk~\cite{LLM,Yang} the microscopic tunneling perturbations allowed by its symmetry and determining their projections onto the domain-wall operators, is a natural next step.

\FloatBarrier
\section{Constraints on Holographic Boundary Spectra}
\label{sec:gravity}
A holographic brane junction must be compatible with a spectrum of boundary operators satisfying crossing symmetry. Our inequalities test this requirement using the boundary entropies and the dimension of the boundary-condition-changing operator inferred from the geometry. In the solutions of Refs.~\cite{MiyajiMurdia,BiswasEtAl}, the junction interaction changes the semiclassical prediction for this dimension while the brane tensions, and hence the boundary entropies at leading order, remain fixed. At fixed unequal tensions, a boundary-condition-changing operator that is light compared with the central charge requires light nonvacuum primaries on the boundary with larger boundary entropy.

\subsection{Euclidean Action and Brane Junctions}
In AdS$_3$/BCFT$_2$, a conformal boundary of a two-dimensional CFT is represented by an end-of-the-world brane on which the three-dimensional bulk spacetime ends~\cite{Takayanagi,FujitaTakayanagiTonni}. Consider the CFT on a strip of width $L$, with boundary conditions $a$ and $b$ at its two ends. Its gravity dual has the strip as its asymptotic boundary $N$, with two branes $Q_a,Q_b$ anchored at the ends of the strip. In the models of Refs.~\cite{MiyajiMurdia,BiswasEtAl}, the branes meet along a line $J=Q_a\cap Q_b$ in the bulk. A localized interaction at $J$ allows branes of different tensions to join.

The Euclidean action for Einstein gravity with branes and a junction interaction is~\cite{MiyajiMurdia,BiswasEtAl}
\begin{equation}
\begin{aligned}
I_E={}&-\frac{1}{16\pi G_N}\int_M\!\sqrt g\,(R+2/\ell^2)
-\frac{1}{8\pi G_N}\int_N\!\sqrt h\,K\\
&-\frac{1}{8\pi G_N}\sum_{i=a,b}\int_{Q_i}\!\sqrt h\,(K-T_i)
-\frac{1}{8\pi G_N}\int_J\!\sqrt\sigma\,(\theta_0-\theta)\\
&-\frac{1}{8\pi G_N}\sum_{i=a,b}\int_{N\cap Q_i}\!\sqrt\sigma\,(\pi-\theta_i)
+I_{\rm ct}.
\end{aligned}
\label{eq:adsaction}
\end{equation}
Here $M$ is the bulk region, $R$ is its Ricci scalar, $\ell$ is the AdS radius, and $G_N$ is Newton's constant. The metric determinants on $M$, its boundary surfaces, and their intersections are denoted by $g$, $h$, and $\sigma$, respectively. The extrinsic curvature $K_{\mu\nu}$ is defined with outward normals, and $K$ is its trace. The brane tensions are $T_i$, with $|\ell T_i|<1$. The angles $\theta$ at $J$ and $\theta_i$ at $N\cap Q_i$ are internal angles of $M$. The last line contains the corner terms at the asymptotic boundary and $I_{\rm ct}$, the local counterterms needed to render the action finite.

The parameter $\theta_0$ controls the interaction proportional to the junction length. At $\theta_0=\pi$, the junction term reduces to the Hayward corner term~\cite{Hayward}. With the metric on $N$ held fixed, the brane and junction conditions are~\cite{MiyajiMurdia,BiswasEtAl}
\begin{equation}
K_{\mu\nu}-(K-T_i)h_{\mu\nu}=0\quad\text{on }Q_i,
\qquad \theta=\theta_0\quad\text{on }J.
\label{eq:adsboundaryconditions}
\end{equation}
Thus each tension fixes the extrinsic curvature of its brane, while the junction interaction fixes the angle at which the two branes meet. At leading semiclassical order, the boundary entropy is fixed by the brane tension and the CFT central charge $c$~\cite{Takayanagi,FujitaTakayanagiTonni}:
\begin{equation}
\log g_i=\frac c6\operatorname{arctanh}(\ell T_i)+o(c),
\qquad c=\frac{3\ell}{2G_N}.
\label{eq:gravity}
\end{equation}

\subsection{Strip Energy and Boundary-condition-changing Operators}
The state--operator correspondence relates states on a strip with boundary conditions $a,b$ to boundary-condition-changing operators~\cite{Cardy,CardyReview}. If $\delta$ is the smallest such dimension, the ground-state energy on the strip is
\begin{equation}
E_{ab}=\frac{\pi}{L}\left(\delta-\frac c{24}\right).
\label{eq:stripenergy}
\end{equation}
In the $a$ channel, the OPE of this boundary-condition-changing operator with its adjoint contains local operators that preserve $a$. Our inequality constrains the dimensions of these operators.

We use the static connected solution of Ref.~\cite{MiyajiMurdia} with no additional conical defect in the bulk. The region between the two branes is a portion of Euclidean thermal AdS. Let $2\alpha_0$ be the angular separation between the branes at the asymptotic boundary, measured in the AdS angular coordinate of period $2\pi$. It is fixed by the tensions and the junction angle through
\begin{equation}
\cos\theta_0=\sqrt{(1-\ell^2T_a^2)(1-\ell^2T_b^2)}\cos(2\alpha_0)-\ell^2T_aT_b.
\label{eq:braneangle}
\end{equation}
On the branch with $0<\theta_0\leq\pi$ compatible with a unitary BCFT, $0<\alpha_0\leq\pi/2$.

Assuming that this connected solution describes the ground state on the strip, its energy determines the lowest dimension of a boundary-condition-changing operator through Eq.~\eqref{eq:stripenergy}, giving~\cite{MiyajiMurdia,BiswasEtAl}
\begin{equation}
\frac{\delta}{c}=\frac1{24}\left[1-\left(\frac{2\alpha_0}{\pi}\right)^2\right]+o(1).
\label{eq:MM}
\end{equation}
At fixed unequal tensions, the leading semiclassical value of $\delta/c$ can be made arbitrarily small by taking $\alpha_0$ close to $\pi/2$ from below. The configuration at $\alpha_0=\pi/2$ is excluded because the junction reaches the asymptotic boundary and one brane disappears.

\subsection{Light Boundary Operators Required by Crossing}
Fix $T_a>T_b$. Equation~\eqref{eq:gravity} gives a difference of order $c$ between the boundary entropies, so $g_b/g_a$ vanishes exponentially as $c\to\infty$. Applying Eq.~\eqref{eq:main} to the boundary-condition-changing operator of lowest dimension then gives
\begin{equation}
H_a\leq2\sqrt2\,\delta\,[1+o(1)].
\label{eq:holo}
\end{equation}
Here $H_a$ is the lowest dimension of a nonvacuum Virasoro primary in the $a$-channel OPE. If $\delta=o(c)$, the boundary with larger boundary entropy must contain a nonvacuum primary of dimension $o(c)$. If $\delta$ remains bounded as $c$ grows, Eq.~\eqref{eq:lightweight} further implies that the relative contribution to $F_a(1/2)$ from primaries of bounded dimension remains bounded away from zero.

For a concrete restriction on the junction geometry, suppose in addition that $H_a\geq c/24+o(c)$, as motivated by the boundary spectrum of pure gravity~\cite{CMW}. Equations~\eqref{eq:holo} and \eqref{eq:MM} then imply
\begin{equation}
\frac{\delta}{c}\geq\frac1{48\sqrt2}+o(1),\qquad
\frac{\alpha_0}{\pi}\leq\frac12\sqrt{1-\frac1{2\sqrt2}}+o(1).
\label{eq:angle}
\end{equation}
At leading order, the upper bound on $\alpha_0/\pi$ is approximately $0.402$. Junctions closer to $\alpha_0=\pi/2$ therefore require primaries below the assumed boundary gap, even though their classical geometry is well defined. Related boundary spectra have been obtained from the fusion transformation of the vacuum block in the bulk channel~\cite{KusukiWei,KusukiAveragedBoundaries}. Our inequality constrains the dimensions of boundary operators without assuming vacuum-block dominance and complements the bounds obtained from the annulus partition function in Ref.~\cite{CMW}.

\section{Discussion}
Our results provide a way to identify an instability without first calculating the dimensions of the operators responsible for it. Two defect entropies and the dimension of a defect-changing operator can already require relevant operators on the defect with larger defect entropy. When symmetry permits the corresponding perturbations, these few data are enough to exclude a proposed stable fixed point.

For impurity order, the key advantage is that data from a single pinning field defect constrain the number of ordered configurations. At the reported numerical inputs, the bounds exclude stable order with $q\geq3$ symmetry-related orientations in $O(2)$ and $O(3)$ and with $q\geq4$ in $O(4)$, whatever the geometry of the orientations, because the inputs to the bound depend only on their number. In $O(3)$, the bound from the third derivative is essential to exclude three-state order, whose stability cannot be determined from calculations for opposite orientations.

The same data bound the lowest dimension of a domain-wall operator over continuous ranges of relative angles. Configurations in which all pairs of orientations make a common angle require a relevant domain-wall operator between any pair with $0<\theta\leq120^\circ$ in $O(3)$ and $0<\theta\leq109.47^\circ$ in $O(4)$, provided the OPE on each pinning field defect contains no relevant singlet, so the direct sum of two such defects is unstable even under perturbations preserving their exchange symmetry. Our direct fuzzy-sphere computation of the domain-wall dimensions on a grid spanning the full angular range is consistent with these bounds; for the three principal exclusions the fitted estimates lie approximately $25$--$35\%$ below the corresponding bounds at the nominal inputs, and the estimates support the instability of three-state order in $O(4)$, which lies beyond the reach of the bounds. The computed dimensions increase with the angle and with $N$, and the extrapolated domain wall between opposite orientations is irrelevant in $O(4)$ and close to the relevance threshold in $O(3)$. Within this class of direct sums, two-state order therefore remains the only candidate for stable impurity order at $N=3,4$, and the numerical estimates support the absence of stable order at $N=2$. Continuum estimates of the defect entropy and the endpoint dimension with controlled uncertainties would establish the exclusions throughout a specified uncertainty region, under the stated defect and symmetry assumptions, and following the renormalization-group flows triggered by the relevant domain-wall operators would determine whether the impurity restores the symmetry or locks onto a single orientation.

In AdS/BCFT, the bounds test whether a classical brane junction is compatible with an assumed boundary spectrum. When the boundary entropies differ by order $c$, a junction corresponding to a boundary-condition-changing operator of dimension $o(c)$ requires nonvacuum primaries of dimension $o(c)$ on the boundary with larger boundary entropy. Such a junction is incompatible with a boundary gap of order $c$, even when its classical geometry is well defined. Combining these bounds with the bootstrap equation for the annulus partition function could further constrain both the spectrum of boundary-condition-changing operators and the local spectrum on each boundary. For a related approach combining annulus and bulk correlator constraints, see Ref.~\cite{MeineriRadhakrishnan}.

The present bounds do not detect the known instabilities of two-state order in the Ising and $O(2)$ models. They also give no constraint when $qg_{\rm pin}<1$, as for two-state order at $N\geq3$, and the exact spectra of two-dimensional parafermion theories show that the domain-wall dimension need not be monotonic in the relative angle, so bounds covering continuous ranges of angles are needed to reach such cases. 

The bounds presented here are not optimized. Even at fixed derivative order, varying the coefficients of the linear functionals, subject to positivity on the allowed conformal blocks, can yield stronger constraints. For example, at the nominal inputs for the \(O(3)\) model, a different choice of coefficients gives a smaller real-valued upper bound on \(q\), although the excluded integer values remain unchanged. Our aim in this work is to obtain simple, universal analytic bounds from a small set of defect data. We therefore leave a systematic optimization of the functionals, and the determination of the strongest bounds attainable within this framework for individual models, to future work.
Including higher derivatives of the crossing equation could further strengthen the bounds, while four-point functions involving several defect-changing operators could constrain the dimensions of operators that do not appear in a single chosen OPE. These extensions could broaden the range of impurity fixed points whose stability can be tested analytically.

\noindent\begin{minipage}{\textwidth}
\vspace{0.5\baselineskip}
\section*{Acknowledgments}
YK is supported by the INAMORI Frontier Program at Kyushu University, JSPS KAKENHI Grant Number 23K20046, MEXT Supporting Pioneering Research through AI for 1,000 Discovery challenges Program (SPReAD) Japan Grant Number JPMXP1726262823, and JST PRESTO Grant Number JPMJPR26A6. WH is supported by JST SPRING, Japan Grant Number JPMJSP2136. We used ChatGPT (OpenAI, GPT-6 astra) and Claude (Anthropic, Opus 5.5) to assist with drafting, writing the numerical codes and checking the manuscript under our direction. The authors are responsible for all content and conclusions.
\end{minipage}

\appendix
\section{Positivity of Conformal Block Expansions}
\label{app:vir}
For a four-point function with cross ratio $0<z<1$ and external primaries of equal dimension, the global block of an intermediate state of dimension $h>0$ has the radial expansion~\cite{HogervorstRychkov}
\begin{equation}
\begin{aligned}
\mathcal{F}_h(z)&=z^h{}_2F_1(h,h;2h;z)\\
&=(4\rho)^h{}_2F_1\!\left(\tfrac12,h;h+\tfrac12;\rho^2\right),\\
\rho&=\frac{z}{(1+\sqrt{1-z})^2}.
\end{aligned}
\label{eq:radial}
\end{equation}
The second line follows from the quadratic transformation of the hypergeometric function. The series in $\rho^2$ has nonnegative coefficients, and $(\rho'/\rho)_{z=1/2}=2\sqrt2$, where primes denote derivatives with respect to $z$. Differentiating therefore gives
\begin{equation}
\mathcal{F}_h'(1/2)\geq2\sqrt2\,h\mathcal{F}_h(1/2).
\label{eq:appendixderivative}
\end{equation}
The global vacuum block is $\mathcal{F}_0=1$. In a unitary boundary CFT, each Virasoro representation decomposes into orthogonal global conformal representations after null states are removed. For the alternating four-point function of a boundary-condition-changing operator and its adjoint, the matrix elements entering each conformal block are paired with their complex conjugates. Positive norms then give nonnegative coefficients in the global block decomposition. The global primaries have dimensions at least $h$, the dimension of the Virasoro primary. Applying Eq.~\eqref{eq:appendixderivative} term by term proves the same estimate for a nonvacuum Virasoro block and gives $\mathcal{F}_0'(1/2)\geq0$ for the Virasoro vacuum block, as required in Eq.~\eqref{eq:lemma}.

\section{Proof of the Bound from the Third Derivative}
\label{app:third}
We prove the inequality for conformal blocks in Eq.~\eqref{eq:thirdlower} and the final inequality in Eq.~\eqref{eq:thirdrational}. We consider the global blocks in Eq.~\eqref{eq:radial} with $h>0$ and external dimension $0<\delta\leq1/6$. Primes denote derivatives with respect to $z$; all conformal blocks and their derivatives are evaluated at $z=1/2$. The Casimir equation gives
\begin{align}
&\mathcal{F}_h'''+6(\sqrt2-4\delta)\mathcal{F}_h''+12(16\delta^2-8\sqrt2\,\delta+\sqrt2)\mathcal{F}_h'\nonumber\\
&\quad=8\bigl[24\delta^2-6(2\sqrt2+1)\delta+3\sqrt2+1+h(h-1)\bigr]\mathcal{F}_h'\nonumber\\
&\qquad+48(\sqrt2-4\delta)h(h-1)\mathcal{F}_h.
\label{eq:appendixcasimir}
\end{align}
The coefficient of $\mathcal{F}_h'$ on the right-hand side decreases with $\delta$. Evaluating it at $\delta=1/6$ and using $h(h-1)\geq-1/4$ gives
\begin{equation}
8\bigl[24\delta^2-6(2\sqrt2+1)\delta+3\sqrt2+1+h(h-1)\bigr]
\geq\frac{10}{3}+8\sqrt2>0.
\end{equation}
Applying $\mathcal{F}_h'\geq2\sqrt2\,h\mathcal{F}_h$ from Appendix~\ref{app:vir} gives a lower bound $P(h)\mathcal{F}_h$ on the left-hand side of Eq.~\eqref{eq:appendixcasimir}, where
\begin{align}
P(h)&=16\sqrt2\,h^3+32(\sqrt2-6\delta)h^2\nonumber\\
&\qquad+32\bigl[12\sqrt2\,\delta^2-3(2+\sqrt2)\delta+3-\sqrt2\bigr]h.
\end{align}
All three coefficients are positive. The quadratic coefficient is at least $32(\sqrt2-1)>0$, while the linear coefficient decreases with $\delta$ and is at least $64-112\sqrt2/3>0$. Since each of $h$, $h^2$, and $h^3$ is at least $\min\{h,h^3\}$ for $h>0$, we obtain
\begin{align}
P(h)&\geq P(1)\min\{h,h^3\}\nonumber\\
&=16\sqrt2\bigl[24\delta^2-6(2\sqrt2+1)\delta+3\sqrt2+1\bigr]\min\{h,h^3\}.
\end{align}
This proves Eq.~\eqref{eq:thirdlower}.

To obtain the simplified bound, we prove
\begin{equation}
\frac{\sqrt2\,\delta(16\delta^2-12\sqrt2\,\delta+3\sqrt2+2)}{24\delta^2-6(2\sqrt2+1)\delta+3\sqrt2+1}
\leq\frac{7\delta}{4(1-2\delta)}.
\label{eq:appendixrational}
\end{equation}
Clearing the positive denominators and cancelling $\delta$ reduces this inequality to
\begin{align}
&7\bigl[24\delta^2-6(2\sqrt2+1)\delta+3\sqrt2+1\bigr]\nonumber\\
&\qquad-4\sqrt2(1-2\delta)(16\delta^2-12\sqrt2\,\delta+3\sqrt2+2)\nonumber\\
&\quad=128\sqrt2\,\delta^3-(24+64\sqrt2)\delta^2+(102-68\sqrt2)\delta+13\sqrt2-17\geq0.
\label{eq:relaxation}
\end{align}
The cubic on the last line has second derivative $768\sqrt2\,\delta-48-128\sqrt2<0$, so it is concave throughout the interval. Its endpoint values are $13\sqrt2-17>0$ and $(13\sqrt2-18)/27>0$. It is therefore positive throughout, proving the estimate.

For the uncertainty analysis in Section~\ref{sec:inputs}, we also show that the left-hand side of Eq.~\eqref{eq:appendixrational} increases with $\delta$. The denominator on the left-hand side of Eq.~\eqref{eq:appendixrational} is positive and decreases with $\delta$. The numerator vanishes at $\delta=0$ and has derivative
\begin{equation}
48\sqrt2\,\delta^2-48\delta+6+2\sqrt2\geq\frac{10\sqrt2}{3}-2>0.
\end{equation}
The numerator is therefore positive and increasing for $\delta>0$, while the denominator is positive and decreasing. The fraction consequently increases with $\delta$ throughout the stated interval.

\section{Boundary Spectra and Crossing in Two-dimensional CFT}
\label{app:examples}
In the Ising CFT with $c=1/2$, take $a$ to be the free boundary and $b$ a fixed boundary. Their boundary $g$-functions are $g_a=1$ and $g_b=1/\sqrt2$, and the boundary-condition-changing operator has dimension $\delta=1/16$. We use the two-point normalization $\sqrt{g_ag_b}=2^{-1/4}$ of Eq.~\eqref{eq:twopointnormalization}. The corresponding Virasoro blocks are
\begin{align}
\mathcal{F}_0(z)&=(1-z)^{-1/8}\sqrt{\frac{1+\sqrt{1-z}}2},\\
\mathcal{F}_{1/2}(z)&=2(1-z)^{-1/8}\sqrt{\frac{1-\sqrt{1-z}}2}.
\end{align}
The conformal block expansions are
\begin{equation}
F_a(z)=g_b\left[\mathcal{F}_0(z)+\tfrac12\mathcal{F}_{1/2}(z)\right],\qquad
F_b(z)=g_a\mathcal{F}_0(z).
\end{equation}
Substitution verifies the crossing equation
\begin{equation}
F_a(z)=\left(\frac{z}{1-z}\right)^{1/8}F_b(1-z).
\end{equation}

More generally, the Cardy boundaries $a=(1,2)$ and $b=(1,1)$ of the diagonal unitary minimal model $\mathcal M(m,m+1)$, with $m\geq3$, have
\begin{equation}
\frac{g_a}{g_b}=2\cos\frac{\pi}{m+1},\quad
\delta=\frac{m-2}{4(m+1)},\quad H_a=\frac{m-1}{m+1}.
\label{eq:minimal}
\end{equation}
The boundary-condition-changing operator has dimension $h_{1,2}$, and the ratio of the boundary $g$-functions follows from the Cardy boundary states~\cite{Cardy,CardyReview}. 
Here $h_{r,s}$ is defined as $h_{r,s}=\frac{((m+1)r-ms)^2-1}{4m(m+1)}$.
The fusion rule $(1,2)\times(1,2)=(1,1)+(1,3)$~\cite{Runkel} allows only one nonvacuum Virasoro primary, of dimension $h_{1,3}$, in its OPE with its adjoint. Since $g_a>g_b$, Eq.~\eqref{eq:weight} requires a nonzero contribution at that dimension, establishing the value of $H_a$ in Eq.~\eqref{eq:minimal}. The bound $(1-g_b/g_a)H_a\leq2\sqrt2\,\delta$ becomes
\begin{equation}
\left[1-\frac1{2\cos(\pi/(m+1))}\right](m-1)\leq\frac{m-2}{\sqrt2}.
\end{equation}
The cases $m=3,4$ follow by direct substitution. For $m\geq5$, the left-hand side is at most $(m-1)/2\leq(m-2)/\sqrt2$.

To check the bounds for the direct sums in Table~\ref{tab:exact_stability}, take $a$ to be the sum of the fixed boundaries and $b$ the corresponding free boundary in each critical or tricritical model. For each fixed boundary, choose a boundary-condition-changing operator of lowest dimension that connects it to the free boundary, and take the normalized sum of these operators. This sum is invariant under spin flip for Ising and cyclic permutations of the colors for Potts. The equal boundary $g$-functions and equal weights make the $h=0$ contribution to the OPE proportional to the sum of the projectors onto the fixed boundaries. The combinations orthogonal to this sum remain in the spectrum but are not invariant under the symmetry. On $b$, the vacuum is the only primary with $h=0$ in each case. In critical Ising, $g_a/g_b=\sqrt2$ and $\delta=1/16$; each fixed boundary has only the Virasoro vacuum representation, and the primary domain-wall operators between the two fixed boundaries have dimension $1/2$. In critical three-state Potts, $g_a/g_b=\sqrt3$ and $\delta=1/8$; the only nonvacuum Virasoro primary on each fixed boundary has dimension $3$, while the primary domain-wall operators between distinct fixed boundaries have dimension $2/3$~\cite{Potts}.

In the tricritical models, the free boundary $b$ is labelled $0$. Tricritical Ising has $g_a/g_b=\sqrt2$ and $\delta=7/16$. Each fixed boundary contains only the Virasoro vacuum representation, and the primary domain-wall operators between the two fixed boundaries have dimension $3/2$~\cite{AffleckTricritical}. Tricritical three-state Potts has $g_a/g_b=\sqrt3$ and $\delta=3/8$. The only nonvacuum Virasoro primary on each fixed boundary has dimension $5$, while the primary domain-wall operators between distinct fixed boundaries have dimension $4/3$~\cite{IinoTricritical}. These spectra establish stability against symmetry-preserving perturbations, since differences of the projectors are not invariant under the symmetry.

The same fusion and positivity argument used above shows that these domain-wall operators appear in the chosen OPEs. In the critical models, the boundary-condition-changing operators between fixed and free boundaries have Kac label $(1,2)$, while their tricritical counterparts have label $(2,1)$ and fusion rule $(2,1)\times(2,1)=(1,1)+(3,1)$. In each model, fusion allows only the vacuum and primaries with the same dimension as the domain-wall operators listed above. Since $g_a>g_b$, Eq.~\eqref{eq:weight} requires a nonzero contribution at that dimension, establishing the exact value of $H_a$ listed in Table~\ref{tab:exact_stability}.

\bibliographystyle{JHEP}
\bibliography{main}
\end{document}